\documentclass[twocolumn,showkeys] {revtex4}
\usepackage[english]{babel} %
\usepackage {hyperref}
\usepackage
 {amsmath,amssymb,caption,fancyhdr,graphicx,lineno,remreset,sidecap,url}

\usepackage{minitoc}

\begin{document}
\title{RELATIVISTIC QUANTUM ECONOPHYSICS -- NEW PARADIGMS IN COMPLEX 
SYSTEMS MODELLING}

\author{V.~Saptsin}
\email{saptsin@sat.poltava.ua}
\affiliation{Kremenchuk State Polytechnical University, Kremenchuk, Ukraine}

\author{V.~Soloviev}
\email{vnsoloviev@rambler.ru}
\affiliation{Cherkassy National University, Cherkassy, Ukraine}


\begin{abstract}
This work deals with the new, relativistic direction in quantum econophysics, within the bounds of which a change of the classical paradigms in mathematical modelling of socio-economic system is offered. 

Classical physics proceeds from the hypothesis that immediate values of all the physical quantities, characterizing system's state, exist and can be accurately measured in principle.

Non-relativistic quantum mechanics does not reject the existence of the immediate values of the classical physical quantities, nevertheless not each of them can be simultaneously measured (the uncertainty principle).

Relativistic quantum mechanics rejects the existence of the immediate values of any physical quantity in principle, and consequently the notion of the system state, including the notion of the wave function, which becomes rigorously nondefinable.

The task of this work consists in econophysical analysis of the conceptual fundamentals and mathematical apparatus of the classical physics, relativity theory, non-relativistic and relativistic quantum mechanics, subject to the historical, psychological and philosophical aspects and modern state of the socio-economic modeling problem. 

We have shown that actually and, virtually, a long time ago, new paradigms of modeling were accepted in the quantum theory, within the bounds of which the notion of the physical quantity operator becomes the primary fundamental conception(operator is a mathematical image of the procedure, the action), description of the system dynamics becomes discrete and approximate in its essence, prediction of the future, even in the rough, is actually impossible when setting aside the aftereffect i.e. the memory.

In consideration of the analysis conducted in the work we suggest new paradigms of the economical-mathematical modeling.
\end{abstract}

\keywords{econophysics, quantum econophysics, quantum mechanics, 
relativistic quantum mechanics, indeterminancy principle, complexity, system 
theory.}



\maketitle
\label{contents}
\tableofcontents



















\section{Introduction}
\label{sec:Intro}

Econophysics is a relatively new interdisciplinary scientific school, which 
tends to develop itself rapidly, having taken its shape and name in late 
90-ies of the XX century \cite{bib001}. According to our estimation the 
number of original works and articles on the Internet, surveys and 
monographs has already exceeded thousands. Moreover respective courses and 
special subjects are being introduced in the high schools of far and near abroad \cite{bib001,bib002,bib003}.

In Western countries young theoretical physicist, who look for the application 
of their knowledge and abilities not only in physical and technical fields, 
are employed by large corporations, banks, holding companies and other 
subjects of national and world financial and economical activity.

In its classical part econophysics is working on the application of 
mathematical apparatus of statistical physics, random systems physics and 
non-linear physical dynamics included, to discover socio-economic phenomena, 
using one or another physical model and giving the appropriate economical 
interpretation to physical notions, variables and parameters 
\cite{bib001,bib002,bib003,bib004,bib005,bib006,bib007,bib008}. 

Though statistical physics can't get along without quantum-mechanical ideas 
and notions in its fundamentals, the main sphere of its interest is the 
macroscopic description of systems with large number of particles, the 
dynamic behavior of which can't be brought to microscopic dynamical 
equations of quantum mechanics figured out for separate particles without use 
of respective statistical postulates \cite{bib009}.

During last years an increasing flow of works was observed, in which 
detailed models of market process participants interactions and 
quantum-mechanical analogies, notions and terminology based on methods of 
describing socio-economic systems are drawn to explain both particular 
peculiarities of modern market dynamics and economic functioning in whole 
(\cite{bib010, bib011, bib012, bib013, bib014, bib015, bib016, bib017, bib018, bib019, bib020, bib021, bib022, bib023, bib024, bib025, bib026, bib027, bib028, bib029, bib030, bib031, bib032, bib033, bib034, bib035, bib036, bib037, bib038, bib039, bib040, bib041, bib042, bib043, bib044, bib045, bib046, bib047, bib123, bib124, bib125, bib126, bib127, bib128} and quoted there literature).

In spite of discord of names and key word combinations -- quantum economics 
\cite{bib010},\cite{bib011},\cite{bib012}, quantum finances \cite{bib013},\cite{bib014}, quantum market games \cite{bib015}, quantum game theory \cite{bib016},\cite{bib017}, quantum evolutionary game theory \cite{bib018}, quantum economic theory \cite{bib019}, quantum econophysics \cite{bib020},\cite{bib021},\cite{bib022}, quantum decision making \cite{bib125,bib126,bib127,bib128} etc., - the emphasis on using the mathematical apparatus, input equations and quantum-mechanical models is the common feature of all the above listed works. 

Schr\"{o}dinger equation for the wave function \cite{bib012}, von Neumann equation for the density matrix \cite{bib021}, secondary quantization for systems with variable number of particles \cite{bib018,bib023}, the last modifications of which were called the ultra-secondary and the ultra-tertiary quantization  \cite{bib010,bib020}, Ising spin model \cite{bib024, bib025, bib026}, Feinman path integrals \cite{bib027,bib028,bib123,bib124} Bose condensation in quantium liquids \cite{bib029,bib030}, operator representation (Heisenberg representation), interation representation \cite{bib023, bib031,bib032,bib033,bib034}, AdS/CFT correspondence in non-linear quantum finance \cite{bib122} etc. are earning the spotlight.

Among the authors, working purposefully and fruitfully in the field of intersection of quantum physics and economics, we can mention Russian academician V.P. Maslov (\cite{bib020,bib029,bib030,bib035,bib036} and the literature quoted there), researchers from distant foreign countries D. Sornette (\cite{bib129} and the literature quoted there), B.E. Baaquie \cite{bib014, bib028}, C. Pedro Goncalves \cite{bib018, bib037, bib038, bib039, bib040, bib041, bib042, bib043}, E. Guevara Hidalgo \cite{bib021, bib044, bib045, bib046, bib047}.

Although the first works, connected with the application of 
quantum-mechanical models to economic phenomena appeared in the early 90-ies 
of the last century \cite{bib017, bib018, bib048}, it can be confidently contended that a new scientific school in the socio-economic systems modeling is being born. Not going beyond the emerged terminology, it will be the most logical to call this school -- what by the way, most of the authors of the afore-mentioned works, including the authors of this investigation, incline to do -- quantum econophysics \cite{bib018,bib020,bib021,bib022}. 

We consider that the appearance of such a scientific direction is caused not 
only by search for the new applications of quantum mechanics mathematical apparatus  and new quantum-mechanical analogies, but also by the evidently shown problems of the socio-economic modeling, which required deep conceptual analysis and philosophical generalization, including probable change of the established mathematical \cite{bib036} and economic \cite{bib049} paradigms. In the opinion of authors, the relativistic aspects in the conceptual fundamentals of the quantum physics and philosophical reasoning of them, including critical analysis of measurement, state, memory, time and space notions not only in physical, but 
also in psycologycal and socio-economic contexts \cite{bib049,bib053} are gaining great significance in the scope of the new quantum direction in econophysics \cite{bib050, bib051, bib052}. The purpose of this work is the well-reasoned exposition of the totality of the above-mentioned issues, which, as far as we can see, must be related to the competence of the special and dedicated section of econophysics -- relativistic quantum econophysics.

\section{About econophysics, quantum econophysics and complex systems}
\label{sec:econoph}

Econophysics, or physical economics, already mentioned as a relatively young scientific school, recently celebrated its tenth anniversary. Of course that doesn't mean that there were no works on the boundary of economics and physics before the econophysics was officially born, howewer the new direction is usually formed only when the certain conditions appear and the necessity to concentrate the scientific forces arises. Quantum econophysics is not an exception. That is why, though the first work according to Gonsales \cite{bib018}, which can be related to the application of quantum mechanical ideas to the economic phenomena, appeared in 1990 \cite{bib048}, we can speak about the birth of the new scientific direction called econophysics only nowadays. 

In short, quantum econophysics currently includes

\newcounter{N1a} 
\begin{list} {\alph{N1a})}{\usecounter{N1a}}
\item adaptation and usage of mathematical apparatus of quantum mechanics in order to model processes in economics (linear operators in the Hilbert space, wave function and the Schr\"{o}dinger equation, density matrix and the von 
Neumann equation, secondary, ultra-secondary and ultra-tertiary quantization 
apparatus, Feinman path integrals etc.);

\item application of quantum-mechanical models and analogies (the Ising spin glass model, evolutionary quantum game model, Bose condensation of 
quantum fluids etc.);

\item application of quantum mechanical ideology (the uncertainty principle, the 
principle of complementarity, other elements of the quantum measure theory, probabilistic interpretation of the system dynamics).
\end{list}

But, in our opinion, complex analysis of the \textit{conceptual } fundamentals of the modern theoretical physics, basic postulates of the systems theory and system analysis, subject to results of the observations and investigations of the real socio-economic processes and systems, is of no less importance for progress in the correct statement and solving problems of mathematical modeling of complex systems. 

In the contemporary comprehension complex systems are the problem in terms 
of formalization nonlinear systems, in the dynamics of which synergetic 
phenomena are observed, instabilities and poor predictability take place; 
the so-called aftereffect and ``long memory'' connected with it act the 
significant part. First of all socio-economical, ecological and other, which 
are similar to them and depict the upper levels of an integrated, organized 
and functioning in a complicated manner matter, can be related to such 
systems.

Using one or another physical analogy in complex systems modeling, or, as it is often shortly said, in modeling the complexity, we must not forget that 
physics is the experimental science first of all (in principle, as any other 
science is). Each physical theory is based solely on experimental facts, and 
its mathematical apparatus and respective mathematical model is just the 
tool, used to describe the results of observations and/or experiments, which 
is always more or less approximate, and usually not the only one.

Models, describing physical processes, and models, which claim to be 
adequately describing socio-economic processes, are on the essentially 
different and in some way opposite levels of hierarchy of models of the 
world around us. If the physical picture of the world, at least in its 
fundamental principles, does not change for about ten billion years, the 
upper (socio-economic) levels of the matter organization are constantly 
getting more complex and develop in time, and in the last decades it happens 
beneath our eyes. As the time is irreversible -- and this experimental fact 
has not been disproved yet, -- all the attempts to model or predict the 
behaviour of socio-economic or other complicated systems using straight 
``physical'' methods can be rather difficult due to the impossibility of the 
strict following one of the basic exact sciences principle -- the principle 
of experiment and observation results reproducibility.

Of course, models of socio-economic systems should not contradict the 
physical and other processes of the lower level running in them; 
nevertheless not all peculiarities of socio-economic systems can be derived 
from their physical qualities (known in the general system theory emergent 
principle). Actually, this statement can be related to any pair from the 
model hierarchy existing on different levels and describing the world around 
us. 

The straight application of physical approaches and respective to them 
mathematical models in description of socio-economic systems is useful; 
though going beyond the bounds of their applicability may lead to paradoxes 
already observed in the history of science. Mechanical determinism, based on 
the classic Newtonian mechanics; heat death of the Universe, following from 
basic thermodynamics postulates; persistent mathematics paradoxes, derived 
from the infinity notion etc. can be related to the number of such 
paradoxes.

\section{Theoretical physics as one of the reality models and mathematics 
as the formalized language of its description}
\label{sec:Theorphys}

21$^{st}$ century is the century of the triumph of the new theoretical 
physics -- relativity theory and quantum mechanics, which explained new 
phenomena, observed both in macro- and micro-world, as well as changed or 
filled well-established physical notions with the new sense. These notions 
were creating the basis of natural sciences, forming respective philosophic 
concepts and ideas in each and every science without exception, including 
the philosophy itself (the so-called metaphysical approach) for ages.

Though new concepts became firmly established, first of all, 
technologically, as a tool in physics, we consider them to be not fully 
realized yet and used in modeling of socio-economic systems and processes 
running in them.

The reasons of it are hidden not only in the lack of sufficient physical and 
mathematical models spectrum, but also in the inertia, in the absence of the 
deeply integrated analysis, concerning classical physics fundamentals, 
relativity theory, quantum mechanics, theoretical and practical economics, 
as well as historical, psychological, social, philosophical and other, 
strictly ``humanitarian'' aspects of the problem.

In connection with all afore-mentioned, the solution of problems in mathematical modeling of complex systems must be sought on the intersection of various scientific schools, including not only mathematics, physics, cybernetics, computer science etc., but traditionally humanitarian disciplines - philosophy, political science, sociology, psychology, linguistics and others as well - probably, it will be effective to get the new ideas from them \cite{bib129}. V.I.Vernadsky gave classical example of such complex approach to the problem of space and time that keeps until the present \cite{bib130}. Synergetics \cite{bib054}, fractal theory \cite{bib055}, chaos theory \cite{bib056}, econophysics \cite{bib001}, quantum informatics \cite{bib057,bib058,bib059}, neuroeconomics \cite{bib060}, p-adic mathematical physics \cite{bib061,bib062,bib063} and others can be named as the examples of new interdisciplinary directions. Quantum econophysics, which is discussed in this work, can also be related to such an interdisciplinary direction, which we consider to have great perspectives \cite{bib018, bib020, bib021, bib022}.

On the one hand, development of physics and mathematics, appearance of electronic calculating machines and, later, computers, which have performed an informational revolution in all fields of human activity without exception, created the illusion of omnipotence of mathematics as the tool of description, modeling and solving any tasks, connected with the intellectual activity. On the other hand it revealed its shortcomings.

Let us take note that mathematics, as one of the languages of reflection and 
description of the surrounding reality, substantially developed in the scope 
of exact sciences, first of all physics and its technical applications, and 
only thereafter it was used to solve more ``humanitarian'' tasks.

But, as it has been already mentioned before, it is necessary to 
approach an application of physical analogies in modeling of the systems of 
``non-physical'' origin, that occupy the highest (in complexity and time of 
the appearance) levels of models of the universe hierarchy with care. 

Mathematics is built on axioms, and one of its peculiarities is its 
determinacy, the ``rigidity'' of the language used. Unlike the usual 
language, mathematics bars from explanations and contexts; both the strength 
and the narrow-mindedness come from it.

An economist, politician and thinker A.A. Bohdanov wrote about it in his "organized science" (tectology) \cite{bib064,bib065}. (Though as a thinker he treated philosophy rather negatively). His ideas are starting to revive only nowadays.

The writer of genius A.N. Tolstoy in his fairy-tale "The Golden Key, or the Adventures of Buratino" described the inadequacy, to say the least, of mathematical language as the tool of reflection of the gorgeousness of surrounding reality in the allegoric way (episodes of the illness of Buratino, "the patient is either alive or dead", and of his method of solving the elementary sum, "2-1=2") \cite{bib066}.

G\"{o}del was the first to prove the boundedness of mathematics as the language, based on the closed system of axioms, in 1931 in his famous incompleteness theorem \cite{bib067,bib068}; though the true meaning of the theorem, including the philosophical one, is getting fully appraised only now \cite{bib069}. 

Though probability theory, as one of the chapters in mathematics, is 
developed to describe uncertainties, it also brings the problem definition 
to the formally deterministic state, bringing the notion of the probability 
of the event (a determined, strictly given number between zero and one) in, 
and gains substantial sense only for the repeated number of phenomena. 
However not every uncertainty, observed in real systems and processes, first 
of all socio-economic ones, can be described with the help of probability 
language. 

We should note that, in our opinion, both in methodological and conceptual 
aspects, the language of discrete mathematics is gaining special meaning in 
description of complex systems. Discrete mathematics is based on the 
application of the algorithmic (discrete) models; it is constructive in 
realization and gives an opportunity to get rid of the number o 
philosophical paradoxes, which take place in the continuous (``infinite'') 
mathematics.

The famous Banach-Tarski paradox is a striking example of such a 
philosophical ``deadlock'' \cite{bib070}. Acceptance of the so-called axiom of selection in the rigorous set theory allows to split the sphere into the finite number of parts so that it will be possible to make up two spheres, 
equivalent to the initial one. Non-acceptance of this axiom does not lead to 
contradictions \cite{bib071}, though it considerably weakens investigation of the continuous abstract structures in analysis, algebra, topology and other branches of the mathematics.

\section{General systems theory -- language and methodology of solving hardly formalizable problems}
\label{sec:gensystheory}

The creation of the general systems theory in 1951 by Bertalanffy \cite{bib072, bib073}, which in modern interpretation includes the systems theory itself, systems analysis as its methodology, and mathematical modeling as the technological tool \cite{bib074, bib075, bib076}, was one of the attempts to go beyond the limits of the circle of tasks, being solved by the classical ``accurate'' mathematics.

The general systems theory, an integral part of which includes mathematical 
tools, does not exist as the theory in the strict mathematical understanding 
of this word. We suppose it was that very peculiarity of the theory, 
Bertalanffy wanted to emphasize, adding the characteristic ``general'' to 
its definition and mentioning, that even in bounds of the common classical 
mechanics mathematically unsolvable problems appear (three-body problem), 
not speaking about more complex systems and more ``advanced'' models of 
modern theoretical physics. 

The general systems theory can be considered as an empirical set of 
logically unprovable principles, concepts and approaches, which are deduced 
from observations of real complex systems, including those that function 
with the human participation, are common for objects of any nature and appear to be useful when conducting observations, investigation, and, above all, when 
solving practical tasks.

There are various definitions of system. As one of the possible working and 
rather ``integrated'' definitions, which take into account ontological, 
gnosiological and dynamical aspects of the ``system'' notion, we can use the 
following one. 

System is the totality of the interacting elements, into which the subject 
divides the object according to some rules, in order to observe, describe, 
examine and, in the end, solve one or another practical task, meanwhile the 
interaction of the elements in the system when functioning causes the new 
quality, which is not peculiar to the separate element of the system. 

This definition could be considered as the free ``integrated'' 
interpretation of definitions \cite{bib075,bib076}, although we should mention, that considering the essence of the general theory, its statements and initial 
definitions must be neither only, nor ``strict'', nor closed, as it in 
itself is one of the systems, and its own principles are applicable to it.

To the most important principles and statements of the general systems theory, 
which determine the gist of the so-called systems approach the following 
must be related:
\newcounter{N2a} 
\begin{list} {\alph{N2a})}{\usecounter{N2a}}
\item discreteness; 
\item hierarchy;
\item emergence;
\item openness. 
\end{list}

In spite of the absence of direct links between general systems theory, 
which is difficult to formalize, and modern theoretical physics, based on 
the usage of rather abstract mathematical models, both of them, being 
different experimentally grounded ways to reflect the real and only world, 
have deep and common roots. 

First of all we should note that from the definition of the system and 
fundamental statements of the general systems theory follows that within the 
bounds of the systematic approach the question about the objective, i.e. 
non-depending on the subject, existence of the world around us is 
insensible. Of course the world exists regardless of us, but its description 
or reflection is subjective, and the ``subject --object'' couple is in 
compliance with system principles the new system, the properties of which 
under the emergence principle cannot boil down neither to the object's 
properties, nor to the subject's ones taken apart. (In quantum mechanics 
such a philosophical problem of a systematic nature appears when analyzing 
the measuring procedure in the couple ``gauge -- measured object''.)

Continuity, based on the hypothesis of the existence of infinity, which is unprovable in its essence (in the rigorous theory of sets, for example Zermelo-Frenkel axiomatic, this is one of the nine axioms \cite{bib077, bib078}), leads to physical paradoxes, and the systems theory discreteness principle, which is being realized and logically developed in discrete mathematics and theory of algorithms, is the most reasonable alternative to continuity and continuous mathematics, based on it. Likely, continuity isn't the necessary link neither in the physical nor in the mathematical description of the reality \cite{bib079,bib080,bib081,bib082,bib083}. 

The continuity of the basic physical quantities, including those of spatial 
coordinates and time -- is merely a hypothesis and is likely to be an 
approximation, which is not always appropriate for the tasks of 
representation the world around us; therefore within the bounds of the 
systematic approach, when realizing its principles sequentially and to the end, 
these quantities must be also considered as discrete. (It should be 
mentioned that the question of discontinuity and continuity of our time and 
space in physics is still controversial.)

The openness of any system is in certain sense the consequence of its 
hierarchy principle, and the actually observed presence of memory 
(aftereffect) and the registration of time as one of the system-forming factors makes it formally open even when from the very beginning of functioning the system is physically isolated. In the latter case the openness is imparted to the system by its history, the full description and registration of which are 
just impossible (setting the history of the system as the totality of 
initial conditions -- as it is done in the classical physics -- is a quite 
narrow and approximate way of its registration)

\section{Hierarchy of conceptions and models in modern theoretical 
physics}
\label{sec:HoerConcepts}

As it has been already mentioned, theoretical physics of the last century 
fundamentally changed the view on the notions of time and space, measuring 
procedures and the achievable accuracy of the results, on the notion of the 
predictability of system's behaviour; it also put a question of the time 
irreversibility problems, paid attention to presence of the aftereffect 
(memory) in real physical processes. 

One of the most important problems, which should be related to the quantum 
econophysics' competence, is in tracing the influence, which was or will be 
exerted by these changes on problem statement in mathematical modeling of 
the socio-economic processes and interpretation of its results. 

Instrumental approach to physics as to the means of prediction of the 
results of the experiments prepared in certain way is working perfectly in 
the physics itself, nevertheless the transfer of its notions and 
mathematical apparatus on systems of the other nature requires obligatory 
and in-depth analysis of its initial conceptions.

We should note that in modern understanding theoretical physics is the 
hierarchy of models of the physical qualities of the substance, starting 
with classical Newtonian mechanics and ending with the general relativity 
theory and modern parts of relativistic quantum micro- and macro- (cosm-) 
theory, each of them having its own special postulates and own domains of 
applicability. In this regard Newton's laws are as much fundamental as the 
quark or superstrings theory, and those connections, which exist between the 
more and less general theories, as often as not are similar to the temporary 
``bridges'', functioning as the ``scaffolding'' on the theory development 
phase, the rigorous and full substantiation of which usually fails. We will 
concern ourselves with analysis of the conceptual states of the most 
important models mentioned above, making digressions to the general systems 
theory and applying to the practices of real complex systems functioning. 

\subsection{Classical physics and its paradigms -- critical analysis}
\label{sec:ClassicalPhys}

In classical physics it is supposed that basic physical quantities can be 
considered as the quantities, accepting a continuous value series and 
existing \textit{regardless }the measuring procedures. Meanwhile:

\newcounter{N3a} 
\begin{list} {\arabic{N3a})}{\usecounter{N3a}}
\item there are \textit{instantaneous} values of physical quantities, describing the state of the system;
\item in principle there are procedures, which allow to measure the instantaneous values of these physical quantities;
\item the influence of the measuring procedure on the value of the physical 
quantity being measured can be made arbitrarily (negligibly) small.
\end{list}

To such quantities (to make it easier we will confine ourselves to 
mechanics) we can relate -- mass of a $m$ particle, distance (position 
vector $\vec {r}$ with the orthogonal coordinates $x,y,z)$, force (vector 
$\vec {f}$ with projections $f_x ,f_y ,f_z $ on the orthogonal axes of 
coordinates), which can change in time (the time $t$ is absolute, 
continuous, physically irreversible and is considered as a parameter). With 
the help of these and other quantities, which are their derivatives 
(velocity vector $\vec {v} = \dot {\vec {r}}$ with coordinates $v_x = \dot 
{x},{\begin{array}{*{20}c}
 \hfill \\
\end{array} }v_y = \dot {y},{\begin{array}{*{20}c}
 \hfill \\
\end{array} }v_z = \dot {z}$, acceleration vector $\ddot {\vec {r}}$, momentum 
$\vec {p} = m\dot {\vec {r}}$ etc.), using appropriate equations, it is 
possible to make an accurate description of the behaviour of any mechanical 
system. 

Mathematical model is created using Euclidean space, in which existence of 
the inertial coordinate system (Newton's first law), an equation of motion, 
formulated:
\begin{equation}
\label{eq1}
m\ddot {\vec {r}} = \vec {f}
\end{equation}
(Newton's second law for a material particle) and Newton's third law:
\begin{equation}
\label{eq2}
\vec {f}_{12} = - \vec {f}_{21} 
\end{equation}
(force $\vec {f}_{12} $, exerted by the particle 1 on the other material 
particle 2, is of the same magnitude and acts as the opposite to the force 
$\vec {f}_{21} $ direction, exerted by the particle 2 on the particle 1).

Differential and integral calculi serve as the mathematical apparatus for 
solving the problems of classical mechanics, time appears to be the 
independent variable, and system state is characterized by coordinates and 
velocities of its material particles in Euclidean space, system dynamics is 
described by differential equations.

In modern physics instead of Newton's equations are used formalisms 
equivalent to them and based on the principle of least action for Lagrangian 
function of the system or on the Hamilton equations \cite{bib084}, though it does not 
change the essence of the concerned problems. 

Even within the bounds of classical physics assumptions  1)-3) concerning 
physical quantities and relevant measuring procedures are approximations and 
must be considered as hypotheses, true only under certain conditions. 

Indeed, if we proceed not from the abstractions, but from the classical measure theory realities, the notion of the physical quantity (and any other one) is inseparably connected with a certain measuring procedure, which also includes the comparison with some kind of a standard. 

As any measuring procedure takes finite time $\Delta t$, it is assumed that 
during all that time values of the measured physical quantity and essential 
standard's characteristics (or the values of the physical quantity relative 
to the standard) do not change.

Is it really like that? If you think about it, is not quite like that, 
strictly speaking it is not like that at all. For example, the length of the 
bar under the temperature oscillation of the component atoms (or, if the bar 
is under the temperature close to the absolute zero, under so-called 
``zero-point'' quantum oscillations unremovable in their essence) is 
constantly changing. 

It means that the value of the measured bar length, attributed to the $t$ 
moment of the procedure finishing, $x\left( t \right)$, is a certain 
functional (in the simplest case it is a mean value) of the $x\left( {t}' 
\right)$ values when ${t}' < t$:
\begin{equation}
\label{eq3}
x\left( t \right) = F\left[ {x\left( {t}' \right)} 
\right];{\begin{array}{*{20}c}
 \hfill \\
\end{array} }t - \Delta t \le {t}' < t.
\end{equation}

Let us conduct a logical analysis of the ratio (\ref{eq3}), staying within the 
bounds of the classical physics and confining ourselves to the simplest 
one-dimensional case (physical quantity characterizing the system -- scalar) 
for an easy operation. 

If a certain value of some physical quantity $x$ or its projection in a 
given coordinate system (it is not necessary for it to be length or one of 
the point's orthogonal coordinates in the one-dimensional consideration) 
\textit{initially} exists, but depends on time, then there can be two possible equation (\ref{eq3}) 
interpretations:

\newcounter{cc1}
\begin{list}{\arabic{cc1})}{\usecounter{cc1}}
\item in truth two essentially different variables $x$ appear on its both sides 
(\ref{eq3}) -- implicitly and \textit{hypothetically} existing (``the immediate one'') $x\left( {t}' 
\right)$ (on the right side) and $x\left( t \right)$ (on the left), which 
was \textit{really }measured (``the integral one''), while $F\left[ {x\left( {t}' \right)} 
\right]$ is an implicitly defined functional of the implicitly defined 
function $x\left( {t}' \right),{\begin{array}{*{20}c}
 \hfill \\
\end{array} }{t}' < t$;

\item in both parts (\ref{eq3}) appear the $x$ variables of the same nature, ``the immediate'' $x\left( t \right)$, in that case (\ref{eq3}) should be considered as the \textit{functional equation} used to evaluate the unknown function $x\left( t \right)$, with the 
$F\left[ {x\left( {t}' \right)} \right]$ functional, which must take into 
account all the system qualities necessary for the $x$ measuring, including 
its memory about its past, defining, in the end, the $x\left( t \right)$ 
function.
\end{list}

Thereby, the assumption about the a priori existence of the accurate 
immediate values of the physical quantities (as any other ones), independent 
of any measuring procedures -- the postulate, on which the classical 
mechanics is based -- is corroborated by no logical arguments, except our 
assumptions and experience, which is deliberately approximate and limited by 
the observations of the systems of a certain type.

And the last remark, according to the Newton's laws the immediate coordinate 
values and system's particles' velocities assignment in a given moment of 
time completely determines the system's future behaviour, which must be 
considered as a paradox, contradicting the common sense -- there is no 
aftereffect i.e. memory in such a system, and this model is hardly able to 
describe the functioning of the vast majority of real complex systems.

\subsection{Non-relativistic quantum mechanics -- experimental facts, 
postulates and consequences}
\label{sec:NonRelatQuantMech}

The facts, found experimentally, which underlie the non-relativistic 
mechanics are the evidence of the following regulations:

\newcounter{N42a} 
\begin{list} {\alph{N42a})}{\usecounter{N42a}}

\item the indeterminancy principle turns up, thus there is no conception of the 
particle path;

\item physical quantities can possess not every value, в частности the spectrum of the permitted values can be discrete;

\item as in the classical physics it is assumed that physical quantities can have immediate values, but not every set of them can be measured 
simultaneously;

\item the eventual influence of the measuring procedure on its result takes place, meanwhile system state becomes indeterminate in a varying degree after the measuring;

\item every system is an open one in its essence, because the wave function, which helps to characterize the system state in quantum mechanics (the existence of this function is postulated), is formally determined and continuous in all the space. 
\end{list}

There are various and virtually equivalent formulae of the fundamental 
quantum mechanical regulations, nevertheless any mathematical formalism used 
must satisfy all the above-listed conditions and results of the experiments 
carried out. 

Unfortunately, unlike the classical, even the non-relativistic quantum 
mechanics is void of visualization and is not corroborated by the ``common'' 
sense, its rather deep research and understanding has been so far the lot of 
theoreticians of physics and relatively limited quarters of the specialists 
in some number of applied areas. Therefore we find it necessary to give one 
of the shortest (which is one of its merits), but not very extended (under 
the historical causes) quantum-mechanical axiomatics \cite{bib052}, giving the 
corresponding commentaries to it and drawing necessary and useful analogies 
with the observation practice and experience of the theoretical 
generalization of complex systems' behaviour.

Before proceeding to the formulae \cite{bib052}, we will stop at more traditional and historically established approach to the exposition of quantum mechanics and note its peculiarities. 

Most of the ``classical'', if we can say so, descriptions of the initial 
quantum-mechanical postulates, including the well-known course of 
theoretical physics by L.D. Landau and E.M. Lifshitz \cite{bib051}, are carried out according to the following scheme:

\newcounter{N42b} 
\begin{list} {\alph{N42b})}{\usecounter{N42b}}

\item uncertainity principle for the values of physical quantities being 
measured;

\item system wave function and superposition principle;

\item physical quantities operators.
\end{list}

Such a scheme has a historical, psychological and logical explanation. The problem, stated before the famous founders and ideologists of the quantum theory (M.Planck (1858-1947), A.Einstein (1979-1955), N.Bohr (1985-1962), E.Schrodinger (1887-1961), Louis de Broglie (1892-1987), W.Heisenberg (1901-1976), W.Pauli (1900-1958), E.Fermi (1901-1954), P.Dirac (1902-1984), M.Born (1882-1970), V.Fock (1898-1974), D.Blochinzev (1908-1979), L.Landau (1908-1968) and others), was not only in the development of the mathematical apparatus, which would explain results of the physical experiments, not only in understanding the qualitatively new ideology, based on the classical school they grew up on, but also in bringing it home to the minds of the physical society.

Under such circumstances (inevitably) the conceptions formulated could not 
help having one foot in the ``old'' classical quantum physics, and the other 
foot -- in the ``new'' one. However such a ``half-hearted'' approach was to 
become a brake on the noncontradictory philosophical interpretation of its 
laws and wide spread occurrence of its conceptions sooner or later. 

As far back as 1974, when studying in the postgraduate course of the 
Lomonosov Moscow State University and preparing a paper, which dealt with 
philosophical problems of the quantum mechanics, one of the authors paid 
attention to the rapid and thoroughgoing nature of the majority of 
discussions, applied to the differences in quantum-mechanical notions and 
phenomena interpretations, done by different scientific schools, 
nevertheless he did not understand their essence.

As we know, the discussions, connected with the problems of interpretation 
of quantum physics, don't abate even now, and not only physics and 
philosophers take part in them, but, voluntarily or not, scientists from the 
other fields get involved, in their attempts to use quantum-mechanical 
notions and analogies (quantum psychology \cite{bib085}, quantum sociology \cite{bib086}, quantum logic \cite{bib087}, \cite{bib088} etc.).

As it has been already noted the approach to expounding the fundamentals of 
quantum mechanics, which established in \cite{bib052}, is not a traditional one. In the foreword to the first edition of this book, its scientific editor academician N.N. Bogolyubov mentioned the following: ``the merit of this book is in the logical and consistent character of the exposition, based on the rules and regulations, formulated in explicit form''.  However, it seems to us that the compact and explicit exposition of rules, which can be also called axioms or postulates, in their logical sequence, without superfluously looking back at classical physics, is exactly what gives an opportunity to look at the conceptual fundamentals of quantum mechanics in a completely different way and make proper conclusions of both physical and philosophical nature.

Six postulates of non-relativistic quantum mechanics, set out below, are the 
lecturing variant of exposition \cite{bib052} (V.D. Krivchenkov, 1970, MSU, physical faculty).

\noindent

\textbf{A1}. According to the \textit{first postulate }any physical quantity $L$ (except time $t$, 
which is not a physical quantity in non-relativistic quantum theory and is 
considered as an independent parameter) is associated with the linear 
Hermitian operator 
$\mathord{\buildrel{\lower3pt\hbox{$\scriptscriptstyle\frown$}}\over {L}} $.

Rules of the juxtaposition are based on the classical expressions for 
physical quantities and formulated in the following way: 

\newcounter{Naxa1}
\begin{itemize}
\item
classical $x,y,z$ coordinates are confronted with the coordinate operators: 
\begin{equation}
\label{eq4}
\mathord{\buildrel{\lower3pt\hbox{$\scriptscriptstyle\frown$}}\over {x}} 
\equiv x \cdot ;{\begin{array}{*{20}c}
 \hfill \\
\end{array} 
}\mathord{\buildrel{\lower3pt\hbox{$\scriptscriptstyle\frown$}}\over {y}} 
\equiv y \cdot ;{\begin{array}{*{20}c}
 \hfill \\
\end{array} 
},\mathord{\buildrel{\lower3pt\hbox{$\scriptscriptstyle\frown$}}\over {z}} 
\equiv z \cdot {\begin{array}{*{20}c}
 \hfill \\
\end{array} } \Rightarrow {\begin{array}{*{20}c}
 \hfill \\
\end{array} }\vec {r}{\begin{array}{*{20}c}
 \hfill \\
\end{array} } \to {\begin{array}{*{20}c}
 \hfill \\
\end{array} }\widehat{\vec {r}} \equiv \vec {r};
\end{equation}
\item
classical momentum projections $p_x ,p_y ,p_z $ are confronted with momentum 
projections operators:
\begin{equation}
\label{eq5}
\mathord{\buildrel{\lower3pt\hbox{$\scriptscriptstyle\frown$}}\over {p}} _x 
\equiv i\hbar \frac{\partial }{\partial x};
\mathord{\buildrel{\lower3pt\hbox{$\scriptscriptstyle\frown$}}\over {p}} _y 
\equiv i\hbar \frac{\partial }{\partial y};
\mathord{\buildrel{\lower3pt\hbox{$\scriptscriptstyle\frown$}}\over {p}} _z 
\equiv i\hbar \frac{\partial }{\partial z};
\to 
\vec {p} \equiv i\hbar \vec {\nabla }
\end{equation}
($i$ is an imaginary unit, $\hbar = 1,0546 \cdot 10^{ - 27}erg \cdot s$ is 
the Planck's constant, in (\ref{eq4}) and (\ref{eq5}) coordinate representation of operators is used and postulated);
\item
arbitrary classical physical quantity $L = L\left( {\vec {p},\vec {r},t} 
\right)$, which is the momentum and coordinate function (and therewith a time 
$t$ function in the general case) is juxtaposed with the operator: 
\begin{equation}
\label{eq6}
\mathord{\buildrel{\lower3pt\hbox{$\scriptscriptstyle\frown$}}\over {L}} 
\equiv L\left( {\widehat{\vec {p}},\vec {r},t} \right).
\end{equation}
\end{itemize}

These rules reflect the so-called \textit{conformity principle}. Thus the classical systems' total energy 
$E = H\left( {\vec {p},\vec {r},t} \right)$ is associated with the systems' 
total energy operator (Hamiltonian):
\begin{equation}
\label{eq7}
\mathord{\buildrel{\lower3pt\hbox{$\scriptscriptstyle\frown$}}\over {H}} = 
H\left( {\widehat{\vec {p}},\vec {r},t} \right).
\end{equation}

As operator expressions (\ref{eq6}) cannot always have clear and definite 
interpretation, additional rules are brought in. 

Thus, for example, physical quantity $xp_x \equiv p_x x$ can be formally 
associated with three different operators: 
$$
x \cdot \mathord{\buildrel{\lower3pt\hbox{$\scriptscriptstyle\frown$}}
\over{p}} _x \equiv i\hbar x\frac{\partial }{\partial x};
{\begin{array}{*{20}c}
  \\
\end{array} }
\mathord{\buildrel{\lower3pt\hbox{$\scriptscriptstyle\frown$}}\over {p}} _x 
x \cdot \equiv i\hbar \frac{\partial }{\partial x}x \cdot ;
$$
\begin{equation}
\label{eq8}
\frac{1}{2}\left( {x \cdot 
\mathord{\buildrel{\lower3pt\hbox{$\scriptscriptstyle\frown$}}\over {p}} _x 
+ \mathord{\buildrel{\lower3pt\hbox{$\scriptscriptstyle\frown$}}\over {p}} 
_x x \cdot } \right) \equiv
 \frac{i\hbar }{2}\left( {x\frac{\partial 
}{\partial x} + \frac{\partial }{\partial x}x \cdot } \right),
\end{equation}
however, only the last one (symmetric) expression is the Hermitian operator 
and, consequently, the operator of the physical quantity $xp_x $. 

If the function $L = L\left( {\vec {p},\vec {r},t} \right)$ is not 
polynomial to variable $\vec {p}$, its formal expansion into the 
multidimentional Taylor series is used. Problems of the convergence of 
infinite operational and functional series and interpretation of them, which 
occur meanwhile, are the subject of a special discussion, and correspondence 
of conducted theoretical calculations to results of the experiment serves as 
the selection criterion for the operator representation.

\noindent

\textbf{A2}. According to the \textit{second postulate} the given physical quantity $L$ can 
possess only eigenvalues $\lambda _i $ of 
its$\mathord{\buildrel{\lower3pt\hbox{$\scriptscriptstyle\frown$}}\over {L}} 
$ operator:
\begin{equation}
\label{eq9_1}
\mathord{\buildrel{\lower3pt\hbox{$\scriptscriptstyle\frown$}}\over {L}} 
\varphi = \lambda \varphi ;{\begin{array}{*{20}c}
 \hfill \\
\end{array} } \Rightarrow {\begin{array}{*{20}c}
 \hfill \\
\end{array} }\lambda _i ,\varphi _i ;{\begin{array}{*{20}c}
 \hfill \\
\end{array} 
}\mathord{\buildrel{\lower3pt\hbox{$\scriptscriptstyle\frown$}}\over {L}} 
\varphi _i \equiv \lambda _i \varphi _i 
\end{equation}
which are always real under the Hermitian character of 
$\mathord{\buildrel{\lower3pt\hbox{$\scriptscriptstyle\frown$}}\over {L}} $ 
(standard $\lambda _i $ eigenvalues and $\varphi _i $ eigenfunctions problem 
for the linear Hermitian operator 
$\mathord{\buildrel{\lower3pt\hbox{$\scriptscriptstyle\frown$}}\over {L}} 
)$.

It arises from the afore-mentioned postulate that, unlike the classical 
physics, not every value of the physical quantity can be allowed; 
particularly even the \textit{quantized} (discrete) spectrum of its values is possible. The 
hydrogen atom energy permitted values spectrum affords an example of a 
discrete spectrum (it is the only mathematical problem in non-relativistic 
quantum mechanics, related to the real system, which can be approximately 
solved).

In the conceptual sense the first and the second postulates of quantum 
mechanics actually give the first corroboration of a thesis, brought forward 
by us, about the primacy of the \textit{procedure} against its \textit{result}, which is diametrically 
opposite to the conception accepted in the classical physics. In the sequel 
we will repeatedly return to this thesis, weighing in with the arguments and 
proofs in its favour. 

The conformity principle can be considered as an illustration of genetic 
aspects, which characterize perpetual historical development of both 
theoretical physics and scientific cognition in whole, including the 
following phases:

\begin{itemize}

\item filling the old formulae and statements with the new meaning;

\item generation of the new formulae and statements as a result of the conflict 
between the new and the old and mutations, which occur at that time;

\item selection of the well-grounded theories among the set of possible ones.
\end{itemize}

We find it important to note this aspect, because attempts to create the 
``single theory of everything'', to find those universal ``fundamentals'', 
which will give the opportunity to explain and band together everything that 
happens in this world for good, occur very often, even on the modern level. 
Such attempts in our opinion have no prospects even in the field of 
fundamental physics, not speaking of the theories, which claim to give the 
comprehensive and timeless description of socio-economics phenomena. 

\textbf{A3}. According to the \textit{third postulate} every physical system state is associated with the normalized wave function $\psi $:

$$
\psi = \psi \left( {x,y,z,t} \right);
\int\limits_{ - \infty }^{ + \infty } {\int\limits_{ - \infty 
}^{ + \infty } {\int\limits_{ - \infty }^{ + \infty } {\psi \ast \psi dxdy} 
dz} } = $$
\begin{equation}
\label{eq9}
= \int\limits_{ - \infty }^{ + \infty } {\int\limits_{ - \infty }^{ + 
\infty } {\int\limits_{ - \infty }^{ + \infty } {\left| \psi \right|^2dxdy} 
dz} } = 1
\end{equation}
(to make it easier we consider the system which consists of one particle, 
and use the coordinate representation of its wave function in compliance 
with the coordinate representation for the physical quantities operators, 
accepted above).

In classical mechanics dimensioning of $3N$ coordinates and $3N$ momentum (or 
velocity) particle projections -- $6N$ phase coordinates, which presumably 
can be approximately evaluated -- for the system, which consists of $N$ 
particles, completely defines the system state.

In quantum mechanics the system state is specified by the wave function, 
which does not allow defining all classical phase system coordinates both 
accurately and simultaneously. Set of the measurements, that allows defining 
of the wave function is called \textit{full}, and for the system, consisting of $N$ particles the number of such measurements is twice as little (not taking 
into consideration purely quantum spin variables) as the number we get, when 
defining the system state in the classical way, i.e. $3N$.

As the wave function is formally defined in whole space even for the single 
particle, than any real quantum-mechanical system is virtually \textit{open}. In order to 
describe such systems (i.e. to take into account system's interaction with 
its surroundings, if it is not deliberately small) the density matrix 
representation is used \cite{bib051}. 

\textbf{A4}. The \textit{fourth postulate} says that mathematical expectation (the mean value) of the $L$ physical quantity with the $\mathord{\buildrel{\lower3pt\hbox{$\scriptscriptstyle\frown$}}\over {L}} $ 
operator, for the system, which is at the state with the wave function $\psi 
\left( {x,y,z,t} \right)$, is defined by the integral: 
$$
 < L > {\begin{array}{*{20}c}
 \hfill \\
\end{array} } = $$
\begin{equation}
\label{eq10}
= \int\limits_{ - \infty }^{ + \infty } {\int\limits_{ - 
\infty }^{ + \infty } {\int\limits_{ - \infty }^{ + \infty } {\psi \ast 
\left( {x,y,z,t} 
\right)\mathord{\buildrel{\lower3pt\hbox{$\scriptscriptstyle\frown$}}\over 
{L}} \psi \left( {x,y,z,t} \right)dxdy} dz} } .
\end{equation} 

It follows from this postulate that the result of any measurement has, 
actually, ambiguous character. (Physical quantity can possess a 
deterministic value as a result of measurement only if $\psi \left( 
{x,y,z,t} \right)$ agrees with one of the eigenfunctions $\varphi _i $ of 
the $\mathord{\buildrel{\lower3pt\hbox{$\scriptscriptstyle\frown$}}\over 
{L}} $ operator.) The $\left| {\psi \left( {x,y,z,t} \right)} 
\right|^2dxdydz$ quantity is interpreted as the probability of the particle detecting in the differential of volume $dxdydz$. The probabilistic nature 
or, to be precise, the \textit{uncertainty} of measurement result, is the fundamental peculiarity of quantum-mechanical systems.

\textbf{A5}. The \textit{fifth postulate} (the Schr\"{o}dinger equation) defines system evolution 
(change of its wave function $\psi )$ in time:

\begin{equation}
\label{eq11}
i \cdot \hbar \frac{\partial \psi }{\partial t} = 
\mathord{\buildrel{\lower3pt\hbox{$\scriptscriptstyle\frown$}}\over {H}} 
\psi 
\end{equation} 
and plays the same part as the Newton's second law in quantum mechanics 
does.

\textbf{A6}. The \textit{sixth postulate} concerns the identical microparticle system and comes to 
the statement, that particles are \textit{indistinguishable} in such a system. The existence of a 
\textit{spin} -- a new, purely quantum (relativistic) variable, and division of all known particles into two types -- \textit{fermions} (antisymmetric wave function, particles with the half-integer spin) and \textit{boson} (symmetric wave function, particles with the integer spin) are also postulated. 

From the sixth postulate follows the existence of the specific quantum 
(exchange) interaction, which is implemented only in the collective of 
identical microparticles and does not have a classical analog. In the 
conceptual aspect this postulate can be considered as an obvious physical 
illustration of one of the fundamental principles in systems analysis -- the 
\textit{emergence} principle.

Briefly, touching upon the issue of mathematical aspects and omitting the 
details, but emphasizing the conceptual moments, six postulates of the 
non-relativistic quantum mechanics can be reformulated in the following way:

\newcounter{N43a} 
\begin{list} {\arabic{N43a})}{\usecounter{N43a}}

\item \textit{Instead} of the classical notion ``physical quantity $L$'' a new fundamental notion 
is being brought in ``operator of the physical quantity 
$\mathord{\buildrel{\lower3pt\hbox{$\scriptscriptstyle\frown$}}\over {L}} 
$''.

\item Possible (permitted) values of the \textit{physical quantity }$L$ are the consequence (the result) of 
solving the eigenvalues $\lambda $ \textit{mathematical} problem for the operator of the physical quantity 
$\mathord{\buildrel{\lower3pt\hbox{$\scriptscriptstyle\frown$}}\over {L}} $:

\[
\mathord{\buildrel{\lower3pt\hbox{$\scriptscriptstyle\frown$}}\over {L}} 
\varphi = \lambda \varphi .
\]

\item For the system performance a new notion is being brought in -- normalized 
wave function $\psi $: 

\[
\int {\psi \ast \psi d\tau } = \int {\left| \psi \right|^2d\tau } = 1.
\]

\item Classical value of the physical quantity $L$ in the state with the 
normalized wave function $\psi $ is associated with a new quantity -- mean 
value of the physical quantity $ < L > $, which is defined by the ratio:

\[
 < L > {\begin{array}{*{20}c}
 \hfill \\
\end{array} } = \int {\psi \ast 
\mathord{\buildrel{\lower3pt\hbox{$\scriptscriptstyle\frown$}}\over {L}} 
\psi d\tau } .
\]

\item System evolution in time is characterized by its normalized wave function 
evolution, which is defined by solving the Schr\"{o}dinger equation:

\[
i \cdot \hbar \frac{\partial \psi }{\partial t} = 
\mathord{\buildrel{\lower3pt\hbox{$\scriptscriptstyle\frown$}}\over {H}} 
\psi .
\]

\item In the identical particles system all particles are indistinguishable.
\end{list}
The postulates of quantum (non-relativistic) mechanics (postulates\textbf{ 
A1}-\textbf{A6}), which were mentioned above and are in certain sense 
analogous to the laws of Newtonian classical mechanics, are that very basis, 
on which all its theoretical apparatus and practical applications are being 
constructed. Thus, using rather elementary calculations, it is possible to 
show, that from the postulates \textbf{A1}-\textbf{A4} follows the fundamental ratio of uncertainties for coordinates and velocities (or momenta):

\begin{equation}
\label{eq12}
\Delta x \cdot \Delta v \ge \frac{\hbar }{2m};{\begin{array}{*{20}c}
 \hfill \\
\end{array} }\left( {\Delta x \cdot \Delta p \ge \frac{\hbar }{2}} \right),
\end{equation}
where $\Delta x$ and $\Delta v$ ($\Delta p)$ represent the root-mean-square 
errors of measuring the $x$ coordinate and $v = \dot {x}$ velocity ($p = 
m\dot {x}$ momentum) of the particle of the $m$ mass. 

From the ratio (\ref{eq12}) five important for the future conceptual conclusions 
follow in turn:

\begin{itemize}

\item neither particle coordinate nor its velocity can have accurate values, 
because when $\Delta x = 0$ the velocity uncertainty $\Delta v$, and 
therefore the velocity itself turns into infinity, and when $\Delta v = 0$ 
particle is totally delocalized, i.e. it can be detected in any point of the 
physical space;

\item there is no notion of the immediate speed as the Newtonian limit: 

\begin{equation}
\label{eq13}
v\left( t \right) = \dot {x}\left( t \right) =
 \mathop {\lim 
}\limits_{\Delta t \to 0} \frac{x\left( t \right) - x\left( {t - \Delta t} 
\right)}{\Delta t};
\end{equation} 

\item classical particle coordinate and velocity, defining its state in the 
classical mechanics in the $t$ moment of time, can be determined only 
approximately, when $\Delta t$ is \textit{finite} and big enough; 

\item in reality there is no continuous classical particle path -- it is a rough 
notion, which is worthwhile only when $\Delta t$ intervals between adjacent 
measurements of the particle's location are big enough;

\item prediction of the particle's behaviour, deliberately approximate, which is 
defined by the pair of classical phase variables ($x\left( t \right),v\left( 
t \right))$, is possible only when taking into account its history, i.e. 
\textit{aftereffect}, since:

$$v\left( t \right) \approx \frac{x\left( t \right) - x\left( {t - \Delta t} 
\right)}{\Delta t} =$$
\begin{equation}
\label{eq14}
= \frac{1}{\Delta t}x\left( t \right) - \frac{1}{\Delta 
t}x\left( {t - \Delta t} \right)
\end{equation} 
depends both on $x\left( t \right)$ and $x\left( {t - \Delta t} \right)$.
\end{itemize}

We can also approach to the conclusion about the presence of aftereffect on 
basis of analysis (\ref{eq14}), from the other side. Juxtaposing the classical 
velocity definition (\ref{eq13}) with the uncertainty ratio (\ref{eq12}) we realize that in (\ref{eq14}) neither $x\left( t \right)$, nor $x\left( {t - \Delta t} \right)$, nor both of these quantities simultaneously can be defined \textit{accurately} (otherwise the accurate value of the limit (\ref{eq13}) would exist too), and the uncertainty depends on $\Delta t$, and when $\Delta t \to 0$ (disappearance of the aftereffect) it formally becomes infinitely large (impossibility of the prediction). 

\textit{Thus, quantum mechanics eliminates the classical mechanics paradox, connected with the absence of the aftereffect in mathematical models, used by it.}

From quantum-mechanical analysis of the system and measuring ``tool'' 
interaction process (analysis, which was based on the perturbation theory) 
also follows that the uncertainty of the system energy value $\Delta E$, 
acquired as a result of such an interaction, is connected with its duration 
$\Delta t$ in the ratio:

\begin{equation}
\label{eq15}
\Delta E \cdot \Delta t\sim \hbar 
\end{equation} 

From quantum-mechanical analysis of the particle momentum measuring 
procedure, taking into account (\ref{eq15}), follows one more ratio, which is useful for the future and connects the minimal possible uncertainty of the $\Delta p$ momentum with the duration of its measuring $\Delta t$ and change of the particle velocity $\Delta v$ during the time of measuring \cite{bib051}:
\begin{equation}
\label{eq16}
\Delta v \cdot \Delta p \cdot \Delta t\sim \hbar .
\end{equation} 

It seems important to us to emphasize one more time, that in the 
quantum-mechanical axiomatics, expounded above, the measuring procedure, not 
values of the physical quantities as it was in the classical physics, moves 
to the first place. Meanwhile, as it follows from the postulates, the result 
of the measurement in the general case has probabilistic nature, not every 
value of the physical quantity can be permitted, and the system state turns 
out to be more or less uncertain, because of the uncontrollable interaction 
between the observed system and measuring tool.

The fact that the existence of immediate values of the physical quantities 
is actually \textit{conceded} in non-relativistic quantum mechanics, allows bringing the wave function or density matrix (for open systems) in as the characteristic of the current system state. Meanwhile the wave function can have various representations (coordinate, momentum, matrix representation in one or 
another total system of proper functions, in state occupation numbers 
within the secondary quantization apparatus etc.).

Though, as it follows from the premises, the analysis, conducted even within 
the non-relativistic quantum mechanics apparatus, is the evidence of the 
idea, that there are neither immediate nor accurate values of the physical 
quantities for \textit{real} systems and \textit{real} measuring procedures. Within the bounds of non-relativistic quantum mechanics existence of the immediate accurate values of the physical quantities is a \textit{hypothesis} useful for theory and practice, but impossible to confirm for sure by logical or experimental conclusions, as in the case of classical mechanics.

Let us make a number of remarks that are important in our opinion.

It is normal to consider non-relativistic quantum mechanics as a linear 
theory (e.g. \cite{bib019}), since the carrier of information on the current system state -- its wave function -- is subject to the linear equation -- the 
Schr\"{o}dinger equation, and physical quantities operators are entirely 
linear operators. Nevertheless it is not quite so. 

The notion of a linear operator or linear transformation includes a 
superposition principle and at least assumes that the set of input and 
output elements form a linear space. But physical sense is peculiar only to 
normalized wave functions, i.e. to the solution of either the 
Schr\"{o}dinger equation or the eigenfunction and operator eigenvalue 
equations with \textit{additional} normalizing conditions. Though the set of \textit{normalized} wave functions belongs to the linear space, it does not form the linear space on its own account.

It is well-known that a great number of non-linear problems exist within the 
bounds of classical mechanics, which is considered as a particular extreme 
case of quantum mechanics. In terms of common sense it seems to strange, how 
the more general and formally linear theory generates frequent non-linear 
problems.

Of course there are no paradoxes in it, and everything falls into place, if 
we take into account that mathematical formalism of quantum mechanics is, 
first of all, the \textit{operator} formalism, based on operation algebra with special commutation relations, which is not linear at all; and the wave function is the secondary mathematical object derived from formalism. 

Generally speaking, nonlinearity, as a concept opposite to the linearity 
notion, is substantial for rather narrow mathematical model class, underlain 
by linear (vector) space. Thus, for example, there is no point in speaking 
about nonlinearity of Boolean algebra and probability theory, they are just 
\textit{different} mathematical models. However, the absolutization of the notion of value of the physical quantity, which is in the essence the natural element of the natural linear space, has lead to the absolutization of the notion of nonlinearity, including appearance of the disorienting and therefore poor, in our opinion, term: ``nonlinear science'' (i.e. science, which differs from the linear one). 

Thereby, here we, conducting historical and logical analysis of notions, 
implicitly find arguments approving thesis about the priority of the 
measuring procedure against its result in quantum mechanics, i.e. value of 
the physical quantity, which can be considered as the characteristics of the 
current system state, which is secondary and deliberately subordinate. 

And the last remark, it is well-known that time in quantum dynamic equations 
(e.g. the Schr\"{o}dinger equation for the wave function in the coordinate 
representation) is formally reversible, but the specific character of 
quantum-mechanical monitoring (measuring) procedures makes it irreversible. 
So long as in the reality time is really irreversible, it will be natural to 
include time irreversibility into the axiom scheme as an experimentally 
found fact. Thereto it is enough to change the emphases, taking the primary 
nature and necessity of the measuring procedure (i.e. the action) as the 
basic, and, naturally, accepting the presence of the aftereffect and the 
influence, which the measuring procedure has on the result. In this case the 
question of the time irreversibility and existence of sets of parameters or 
variables, with the help of which it is possible to describe the system 
state and its evolution in time accurately, within the bounds of properly 
formulated axiom scheme of non-relativistic quantum mechanics, loses its 
philosophical currency.

Thus, we think that even on the ground of the analysis of non-relativistic 
quantum mechanics we have reason to accept the afore-mentioned hypothesis 
(thesis) which states that the notion of state in quantum physics is 
neither primary, nor fundamental. If we consider this very hypothesis as a 
``bridge'' and rely on the emergence principle, it will be reasonable to 
found any theory of sufficiently complex dynamic systems on it. 

The actual proof of this hypothesis can be found, through analyzing the real 
dynamics of real systems of any nature. However the most valid arguments for 
such conclusions, and they are just from natural sciences (which is 
extremely important from historical, psychological and philosophical points 
of view), are given by relativistic quantum physics. So let us proceed to 
the analysis of its conceptions.

\subsection{Relativistic quantum mechanics. New paradigms in complex system 
modeling }
\label{sec:RelatQuantMechNewP}

Relativistic quantum mechanics is considered to be not entirely complete yet  because of the lack of the proper experimental basis.

We should note that full experimental substantiation of one or another modern relativistic theory requires energies up to $10^{20}ev$  and more, which are yet inaccessible under terrestrial conditions (particles with such energy are relatively seldom registered with the help of extensive air shower method in cosmic rays),  though some of the problems can be solved by the recently launched collider \cite{bib089}, which is able to give the interaction energy up to $14 \cdot 10^{12}ev$ on the colliding electron beams).

Nevertheless, the results, already achieved within its bounds of relativistic physics (and achieved rather long ago), corroborate the analysis conducted above and its conclusions, giving it not only technical but also conceptual character. 

Among the new statements of relativistic quantum mechanics is the 
fundamental one, which says that any measuring procedure takes 
\textit{fundamentally finite} time $\Delta t$, therefore \textit{there are no} immediate values of physical quantities. The limiting error (terminologically, we think, it will be more accurately to say, limiting uncertainty) of measuring any physical quantity is in this case increasing with the decrease of the time of measuring and finite under any finite $\Delta t$, and the value itself can be attributed only to this time interval $\Delta t$ \cite{bib050}.

Thus, if taking into account the relativistic constraint on the maximum 
possible change of velocity $\Delta v\sim c$ ($c$ represents the light 
speed) in the ratio (\ref{eq16}), it is possible to get the \textit{relativistic} quantum uncertainty 
principle, expressed by the ratio \cite{bib050}:

\begin{equation}
\label{eq17}
\Delta p \cdot \Delta t\sim \hbar / c.
\end{equation} 

Thus the accurate value of a particle momentum can be obtained only when the 
time of measuring is equal to infinity, and it means that only one free 
particle momentum can be accurately measured, when the particle is in such 
(free) state for an infinite amount time.

We should note that mathematical formalism based on the Lie groups and algebra (algebra of operators, which follow certain commutation relations \cite{bib094}) is used to construct the vast majority of modern models in relativistic quantum 
mechanics, including the latest theories \cite{bib090}, \cite{bib091}, \cite{bib092}, \cite{bib093}. Thereby in mathematical formalism of relativistic quantum mechanics the dominating part of the procedure, of the action, is in fact "legitimated", and operator is its formal representation or mathematical image.  

As we have already mentioned, non-relativistic quantum mechanics is created 
on the possibility of immediate measuring of quantities, which characterize 
the system, in principle. Just that very assumption gives an opportunity to 
bring the wave function in as the means of an unbiased description of system 
state and its evolution in time \cite{bib050}, and consequently the notion of state as the fundamental system characteristic can be brought in as well. 

Within the bounds of relativistic quantum mechanics this assumption is 
rejected, therefore the so-called scattering matrix or S-matrix gains in the 
biggest importance. This matrix allows, if the noninteracting particle 
system (when $t = - \infty )$ states are known, predicting the probability 
of various free particle system states, which occur after the interaction, 
when $t \to + \infty $ \cite{bib050}.

Such a ``refined'' statement of the problem of experimental investigation of 
relativistic quantum effects can hardly correspond with the overwhelming 
majority of real physical processes which occur in nature, though it helps 
to get rather accurate and reproducible results and is rather useful for the 
elucidation of fundamental, but only physical laws of nature.

As far as we know, relativistic effects in quantum econophysics in the 
aspects that were touched upon above have not been discussed till now. 
However it doesn't mean that there are no analogues of relativistic effects 
or their consequences in socio-economic processes, so long as the quantity, 
playing the part of the maximum possible velocity in these processes, 
doesn't have to be connected with the physical light speed $c$.

Thereby in terms of conceptual statements of relativistic quantum mechanics, 
taking the conducted analysis into account we have all reasons for accepting 
the hypothesis, which states that the particle measuring procedures (applied 
to \textit{any type} of particles) take finite time in socio-economic systems as well, and the results of measurement depend on the chosen procedure and are secondary against the latter. It is also reasonable to accept the hypothesis, which says that there are no immediate values of economical and other quantities 
and indices, and the accuracy of measuring decreases when the time of the 
measuring diminishes (or these quantities lose their primary sense completely). The latter can be interpreted as one of the corroborations for the hypothesis of occurrence of the non-excludable aftereffect in system, i.e. memory. 

Let us proceed to the further ``relativistic'' conclusions. The ultimate 
accuracy of measuring increases with the increase of its duration, but it is 
possible only when the system is in the constant state; therefore there is 
an \textit{optimal} time of the measuring for real dynamic systems, which means that the optimal observation (measurement) on the system presently must be of 
\textit{discrete} nature in time. The stride parameter, of course, depends on \textit{what }is measured, and on the way \textit{how} it is measured, and optimality has a subjective component in certain sense.

Formally (and not only formally in our opinion) any calculations in 
socio-economic systems, that involve the totality of initial data, including 
the dynamics prediction, must be labeled as complex \textit{indirect} measurements (observations) in compliance with some kind of an algorithm. Thereby the algorithm becomes a \textit{measuring procedure}, which generates the quantity respective to it, while the realization of this procedure, as the realization of any other one, can change the system state and its future behaviour unpredictably. 

To make our conclusions even more convincing, we will carry out the 
following mental experiment. Let us assume that some kind of an 
authoritative and personally uninterested higher being (let us call him ``SSS'' - the ``supercomputer'' with ``super-memory'' and ``supermodel''), having all 
the information on our world (including the information on its history), 
able to conduct any calculations and predict the future arbitrarily 
accurate, is predicting the rise of the dollar/euro cross-rate every other 
month roughly at 10{\%} (dollar tumble).

If this information is inaccessible for others, it is likely to be like 
that. Not much will change in predictions, if ``SSS'' brings this 
information home to one of the businessmen, smart enough to carry out proper 
banking operations without any fuss and increase his capital every other 
month. If \textit{everybody} gets this authoritative information, which is beyond any doubt 
(it comes from ``SSS'' himself!), the dollar/euro cross-rate will rise not 
every other month but every other day, and not at 10{\%} but at dozen 
percents, if not times. 

Let us assume that ``SSS'', having imparted the first variant of his 
forecast to all the interested participants, will consider the other variant 
of it, taking into account that everybody is acquainted with his first 
version (which can be interpreted as expectations now) and has already made 
a decision. If this variant is known to the public at large as well, 
everything will be repeated all over again. 

Within the bounds of the hypothesis of continuous time and infinite 
(``untimely'') computation velocity, such ``ping-pong'' between ``SSS'' and 
users of his information can go on endlessly, what leads to the insoluble 
paradox of both prognostication and real behaviour of the socio-economic 
dynamics.

Within the bounds of the foregoing approach, if we reject the infinity (and, of course, continuity) as the conceptual notion, such paradox simply will not appear within the bounds of the hereinabove explained approach - any "ping-pong" takes time, and , if this "future" will have become the "past" by the moment of prediction, the predictions will become pointless (In this regard the real observed dynamics of the real world can be interpreted as the  real-time work of some kind of  the "utmost", unique and inimitable gigantic "supercomputer", when it is of no importance whether it is a determinate one or has some uncertainties. Our world is virtually such a one.) 

We should mention that the part of not abstract ``SSS'' can be played by the 
possessor of a prediction technology (which is unique and rapid enough for 
that time) who has the necessary information content. It is he, who, being 
personally interested, can gain the local, in time or other financial and 
economical ``coordinates'', profit.

It is evident that the abilities of such a materialized ``SSS'' depend on 
the historical experience, accumulated by this civilization, and mastered 
mass prediction technologies, therefore the socio-economic dynamics and 
reality of the ancient world, the Middle Ages, these days and of the more or 
less distant future -- are different in their essence.

Thus, the new paradigms arise from our analysis, which is considerably based 
on the conceptual fundamentals of relativistic quantum mechanics. On our 
opinion, these paradigms must be accepted and taken as a principle of 
mathematical modeling of complex systems. In expanded form these conceptual 
statements can be formulated in the following way:

\begin{itemize}

\item Priority of the measuring procedure against its result and its unavoidable 
influence on it;
\item Absence of the notion of immediate value of the physical quantity as a 
matter of principle, and, consequently, absence of the notion of system 
state as its fundamental characteristics;
\item Discretness and approximate nature of the system time dynamics (the dynamics is considered as the sequence of system definitionally approximate states under review); 
\item Presence of the irremovable aftereffect, i.e. memory;
\item Finite length and influence of any measuring procedure, including observation and prediction, derived from realization of the algorithmic procedure, on system state and its future behaviour;
\item Refusal of the infinity as the conceptual notion;
\item Time irreversibility.
\end{itemize}

\section{Algorithmic models with discrete time}
\label{sec:AlgDiscrTime}

The statements, expounded above, seem quite obvious to us not only in terms 
of physics, but with a view to the observation practice, research and real 
functioning of socio-economics systems. Therefore they must be taken into 
account during the mathematical statement of relevant problems. 

Algorithmic models are gaining importance in connection with it, being 
discrete in their essence and putting the algorithm, i.e. the procedure, the 
action, with the help of which one or another process fulfils itself, on the 
first place.

It is well-known that the algorithmic approach, was developed in due time by 
A.N. Kolmogorov (the Kolmogorov complexity theory, 1956) \cite{bib095}, who foreknew the great future for it. It was he who made one of the first indications of the priority and independence of the discrete approach (against the continuous one) in the modeling of complex systems \cite{bib079}.

It seems to us that within the bounds of this very approach, when using the 
algorithmic models extensively, we can take into consideration and implement 
all the above-listed conceptual statements, concerning the problem 
definition and solving in mathematical modeling of complex systems. Let us 
consider one of such opportunities.

\subsection{General statement of the discrete modeling problem}
\label{sec:GenStatem}

Sufficiently great algorithm class of models with discrete time can be 
specified by the recurrent process of the following form:

\begin{equation}
\label{eq18}
\vec {x}_{n + 1} = \vec {f}_n \left( {\vec {f}_{n - 1} \left( {...\left( 
{\vec {f}_0 \left( {\vec {x}_0 } \right)} \right)} \right)} 
\right),{\begin{array}{*{20}c}
 \hfill \\
\end{array} }n = 0,1,2,...,
\end{equation} 
where $\vec {f}_i (\vec {x}_i )$ is representing the nonlinear mapping of 
the multidimensional vector $\vec {x}_i $, $i$ is for discrete, real or 
fictitious time, $\vec {x}_0 $ is an input pattern which is considered to be 
set in every member function. In the particular case it is possible for 
$\vec {f}_i (\vec {x})$ not to depend on discrete time $i$, $\vec {f}_i 
(\vec {x}) \equiv \vec {f}(\vec {x})$ (autonomous models):

\begin{equation}
\label{eq19}
\vec {x}_{n + 1} = \vec {f}\left( {\vec {x}_n } 
\right),{\begin{array}{*{20}c}
 \hfill \\
\end{array} }
n = 0,1,2,....
\end{equation} 

Autonomous models usually describe systems, which are considered to be 
isolated. Strictly speaking, the process (\ref{eq18}) can be considered a recurrent 
one only in this case, although formally process (\ref{eq19}) can be made 
autonomous, by giving discrete time the dependent variable status $n \equiv 
y_n $ and adding the ratio $y_{n + 1} = y_n + 1$ (but in this case the new 
process, formally autonomous, will have the deliberately unlimited 
amplitude).

Within the bounds of the model (\ref{eq19}) we will be interested in divergent, 
limited, nonperiodical sequences, since they can reflect complex processes, 
occurring in real systems without the participation of exogenous (external) 
factors.

Determinate chaos models \cite{bib054, bib056, bib096}, neural networks \cite{bib097,bib098,bib099} and continuous models, based on differential and integral equations (after being realized in one or another difference scheme \cite{bib100}) virtually come to the models (\ref{eq18}), (\ref{eq19}).

However, on our opinion, classical differential and integral equations form 
a rather narrow model class, which does not involve all the problem spectrum 
of modern complex system theory, since, as it has been already mentioned, 
differential equation don't include aftereffect and the integral ones don't 
take into account all possible nonlinearities, that can occur in the system 
(the integrating operation is linear by definition). In addition, both of 
them are based on the untestable hypothesis of the existence of infinities 
and assume the existence of limits, which not always takes place.

Identification of the model (\ref{eq18}) comes to the function $\vec {f}_i (\vec 
{x}_i )$ definition, and the differences between determinate chaos models 
and neural networks are connected with the form and methods of defining 
these functions (in the neural network models a narrow, from the 
mathematical point of view, representation class $f_i (x_i )$ is used). 
Generally speaking steadiness or convergence of the processes (\ref{eq18}), (\ref{eq19}) is not assumed, and either a single-stage $x_i $ vector component set or their time history can be of interest.

A single-component model with the memory of the following form:

\begin{equation}
\label{eq20}
x_{n + 1} = f\left( {x_n ;x_{n - 1} ;x_{n - 2} ;...x_{n - k} } 
\right);{\begin{array}{*{20}c}
 \hfill \\
\end{array} }k \ge 1
\end{equation} 
can be also brought down to the model (\ref{eq19}) relative to the $(k + 
1)$-dimensional vector $(x_n ;y_n^{(1)} ;y_n^{(2)} ;...y_n^{(k)} )$, when 
the proper lag variables are being brought in:

\[
y_n^{(1)} = x_{n - 1} ;{\begin{array}{*{20}c}
 \hfill \\
\end{array} }y_n^{(2)} = x_{n - 2} ;{\begin{array}{*{20}c}
 \hfill \\
\end{array} }...{\begin{array}{*{20}c}
 \hfill \\
\end{array} }y_n^{(k)} = x_{n - k} .
\]

Thus, due to the finite time digitization models with memory can be created 
on basis of the model (\ref{eq19}), though it does not contain aftereffect (the 
future depends only on the present).

The question on, whether it is possible to bring the vector model without 
memory (\ref{eq19}) with $(k + 1)$ components to the model with memory (\ref{eq20}) for one 
of the components (this procedure has a certain analogy with the process of 
combining a system of the first-order differential equations into a single 
one of the higher order), requires separate consideration, which will be 
carried out later.
\subsection{On the time irreversibility. The Verhulst model}
\label{sec:AftereffectLongMem}

Within the bounds of the model (\ref{eq19}) time irreversibility can be concerned as 
the biunique correspondence between the vectors $\vec {x}_n $ and $\vec 
{x}_{n + 1} $ on the certain subset $X_i $ of the system (\ref{eq19}) phase space $X$:
$$\vec {x}_{n + 1} = \vec {f}\left( {\vec {x}_n } 
\right);{\begin{array}{*{20}c}
 \hfill \\
\end{array} }\vec {x}_n = \vec {f}^{ - 1}\left( {\vec {x}_{n + 1} } 
\right);{\begin{array}{*{20}c}
 \hfill \\
\end{array} }$$
\begin{equation}
\label{eq21}
\vec {x}_n ,\vec {x}_{n + 1} \in X_i \subseteq X.
\end{equation} 

In the general case $X_i $ must include the system attractor -- the subset 
$X_a $, and belong to the subset $X_0 $, which is a subset of initial 
values, drawing the system up to the attractor $X_a $:

\begin{equation}
\label{eq22}
X_a \subseteq X_i \subseteq X_0 \subseteq X.
\end{equation} 

Let us consider the Verhulst model \cite{bib101,bib102,bib103} as the simplest example. The model is a nonlinear logical and single-component mapping in the following form:
$$x_{n + 1} = f\left( {x_n } \right) = x_n \left( {1 + \alpha \left( {1 - x_n 
} \right)} \right);{\begin{array}{*{20}c}
\end{array} }$$
\begin{equation}
\label{eq23}
0 < \alpha < 3;{\begin{array}{*{20}c}
 \hfill \\
\end{array} }x_0 \in \left( {0;\frac{1 + \alpha }{\alpha }} \right) = X_0 ,
\end{equation} 
where $\alpha $ is a given numerical parameter. We chose the limits for 
$\alpha $ and $x_0 $ so that $x_n $ values would stay positive with any 
chosen $n > 0$.

The largest extremum $x_{n + 1} = x_{\max } $ of the function $x_{n + 1} = 
f\left( {x_n } \right)$ is reached in the point where $x_n = \bar {x}$:

\begin{equation}
\label{eq24}
\bar {x} = \frac{1 + \alpha }{2\alpha };{\begin{array}{*{20}c}
 \hfill \\
\end{array} }x_{\max } = \frac{\left( {1 + \alpha } \right)^2}{4\alpha }.
\end{equation} 
The inverse mapping $x_n = f^{ - 1}\left( {x_{n + 1} } \right)$ is:

$$x_n = \frac{1 + \alpha }{2\alpha }\pm \sqrt {\frac{\left( {1 + \alpha } 
\right)^2}{4\alpha ^2} - \frac{x_{n + 1} }{\alpha }} ;{\begin{array}{*{20}c}
\end{array} }$$
\begin{equation}
\label{eq25}
x_{n + 1} \in \left( {0;\frac{\left( {1 + \alpha } 
\right)^2}{4\alpha }} \right) = X_i \subseteq X_0 
\end{equation} 
and is a two-digit one, generally speaking.

Thereby, the Verhulst model is the one with the irreversible discrete time. 
However, if the following condition is fulfilled:

\begin{equation}
\label{eq26}
x_{\max } \le \bar {x};{\begin{array}{*{20}c}
 \hfill \\
\end{array} } \Rightarrow {\begin{array}{*{20}c}
 \hfill \\
\end{array} }\frac{\left( {1 + \alpha } \right)^2}{4\alpha } \le \frac{1 + 
\alpha }{2\alpha };{\begin{array}{*{20}c}
 \hfill \\
\end{array} } \Rightarrow {\begin{array}{*{20}c}
 \hfill \\
\end{array} }\alpha \le 1,
\end{equation} 
and the interval $\left( {0;\bar {x}} \right)$ is chosen for the $X_i $ 
subset, the inverse mapping becomes a single-digit one.

\subsection{Aftereffect and ``long'' memory in discrete models with 
nonlinearities}
\label{sec:AftereffectLongMemDiscr}

Let us consider the problem of bringing the vector model (\ref{eq19}) to the scalar 
model (\ref{eq20}) for one of the components. Let us start from the case of a 
two-component model:

\begin{equation}
\label{eq27}
\left\{ {{\begin{array}{*{20}c}
 {x_{n + 1} = f_x \left( {x_n ,y_n } \right);} \hfill \\
 {y_{n + 1} = f_y \left( {x_n ,y_n } \right);} \hfill \\
\end{array} }} \right.{\begin{array}{*{20}c}
 \hfill \\
\end{array} }n = 0,1,...
\end{equation}

In order to exclude the $y_i $ variables, we will write down a system of 
three equations for 5 variables $x_n ,y_n ,x_{n + 1} ,y_{n + 1} ,x_{n + 2} 
$, having temporarily equated $n = 0$ to simplify the notation:

\begin{equation}
\label{eq28}
\left\{ {{\begin{array}{*{20}c}
 {x_2 = f_x \left( {x_1 ,y_1 } \right);} \hfill \\
 {x_1 = f_x \left( {x_0 ,y_0 } \right);} \hfill \\
 {y_1 = f_y \left( {x_0 ,y_0 } \right).} \hfill \\
\end{array} }} \right.
\end{equation} 
Let us assume that the second equation of the system (\ref{eq28}) can be definitely 
solved relative to the $y_0 $ variable, i.e. the function $x_1 = f_x \left( 
{x_0 ,y_0 } \right)$ has an inverse one relative to this variable:

\begin{equation}
\label{eq29}
y_0 = f_{x_0 }^{ - 1} \left( {x_0 ,x_1 } \right).
\end{equation} 
Substituting the third equation of the system (\ref{eq28}) into its first one:
$$x_2 = f_x \left( {x_1 ,y_1 } \right)$$
\begin{equation}
\label{eq30}
 = f_x \left( {x_1 ,f_y \left( {x_0 ,y_0 
} \right)} \right) \equiv \tilde {f}_x \left( {x_1 ,x_0 ,y_0 } \right)
\end{equation} 
and substituting the expression (\ref{eq29}) for $y_0 $ in the (\ref{eq30}), we get: 
$$x_2 = \tilde {f}_x \left( {x_1 ,x_0 ,y_0 } \right) =$$
\begin{equation}
\label{eq31}
= \tilde {f}_x \left( 
{x_1 ,x_0 ,f_{x_0 }^{ - 1} \left( {x_0 ,x_1 } \right)} \right) \equiv F_x 
\left( {x_1 ,x_0 } \right).
\end{equation} 

Such memory, the length of which is determined by the number of components 
in the initial vector model (\ref{eq27}) (where the aftereffect is absent), can be called short for convenience.

If the inverse mapping (\ref{eq29}) in the phase variables domain of variation is ambiguous, for example it has two branches:

\begin{equation}
\label{eq32}
y_0 = f_{1x_0 }^{ - 1} \left( {x_0 ,x_1 } \right);{\begin{array}{*{20}c}
 \hfill \\
\end{array} }y_0 = f_{2x_0 }^{ - 1} \left( {x_0 ,x_1 } \right),
\end{equation} 
we should choose the branch, corresponding with the $y_0 $ value, which is 
observed (given) within the initial model (\ref{eq27}), for this pair of variables. 

Thereby the mapping (\ref{eq31}) becomes virtually not only the function of $x_0 
,x_1 $, but also of $y_0 $:

\begin{equation}
\label{eq33}
x_2 = \tilde {F}_x \left( {x_1 ,x_0 ,y_0 } \right).
\end{equation} 
Similarly:

\begin{equation}
\label{eq34}
\begin{array}{l}
 x_3 = \tilde {F}_x \left( {x_2 ,x_1 ,y_1 } \right) = \tilde {F}_x \left( 
{x_2 ,x_1 ,f_y \left( {x_0 ,y_0 } \right)} \right) \equiv \\ \equiv \tilde {\tilde 
{F}}_x \left( {x_2 ,x_1 ,x_0 ,y_0 } \right); \\ 
 x_4 = \tilde {\tilde {F}}_x \left( {x_3 ,x_2 ,x_1 ,y_1 } \right) = \tilde 
{\tilde {F}}_x \left( {x_3 ,x_2 ,x_1 ,f_y \left( {x_0 ,y_0 } \right)} 
\right) \equiv \\ \equiv \tilde {\tilde {\tilde {F}}}_x \left( {x_3 ,x_2 ,x_1 ,x_0 
,y_0 } \right); \\ 
 ... \\ 
 \end{array}
\end{equation} 

It follows from the received correlation chain, that even in the 
two-component system (\ref{eq27}) the ``long'' single-component memory, determined by nonlinear and obligatory nonmonotonic interactions of the components, is actually \textit{possible}. Of course, everything afore-mentioned can be considered to be merely \textit{necessary} conditions for the realization of arbitrary ``long'' single-component memory in systems (\ref{eq19}) with the limited quantity of components; however the wealth of trajectories and phase portraits, observed for such systems during numerical experiments, leaves us hoping for the existence of \textit{sufficient} conditions. To reach these conditions a model with more than two components will be, probably, required, however it does not change the essence of the analysis conducted and conclusions made. The ternary nonlinear Lorenz's mapping \cite{bib104}  can be considered to be one of the examples of the model, where it is possible to realize the ``long'' single-component memory.

Let us briefly consider the scheme of reasoning and computations for the 
ternary model ($N = 3)$:
\begin{equation}
\label{eq35}
\left\{ {{\begin{array}{*{20}c}
 {x_{n + 1} = f_x \left( {x_n ,y_n ,z_n } \right);} \hfill \\
 {y_{n + 1} = f_y \left( {x_n ,y_n ,z_n } \right);} \hfill \\
 {y_{n + 1} = f_y \left( {x_n ,y_n ,z_n } \right);} \hfill \\
\end{array} }} \right.{\begin{array}{*{20}c}
 \hfill \\
\end{array} }n = 0,1,...
\end{equation} 
We will write a set of $k$ equations,

\begin{equation}
\label{eq36}
k = N\left( {N - 1} \right) + 1 = 3\left( {3 - 2} \right) + 1 = 7,
\end{equation} 
for $p$ variables,

\begin{equation}
\label{eq37}
p = N^2 + 1 = 3^2 + 1 = 10,
\end{equation} 
lettered as $x_0 ,y_0 ,z_0 ,x_1 ,y_1 ,z_1 ,x_2 ,y_2 ,z_2 ,z_3 $, having 
equated $n = 0$ to simplify the notation as before (when $N = 2)$:

\begin{equation}
\label{eq38}
x_3 = f_x \left( {x_2 ,y_2 ,z_2 } \right);
\end{equation} 
\begin{equation}
\label{eq39}
x_2 = f_x \left( {x_1 ,y_1 ,z_1 } \right) \equiv f_x \left( {\vec {r}_1 } 
\right);
\end{equation} 
\begin{equation}
\label{eq40}
y_2 = f_y \left( {x_1 ,y_1 ,z_1 } \right) \equiv f_y \left( {\vec {r}_1 } 
\right);
\end{equation} 
\begin{equation}
\label{eq41}
z_2 = f_z \left( {x_1 ,y_1 ,z_1 } \right) \equiv f_z \left( {\vec {r}_1 } 
\right);
\end{equation} 
\begin{equation}
\label{eq42}
x_1 = f_x \left( {x_0 ,y_0 ,z_0 } \right) \equiv f_x \left( {\vec {r}_0 } 
\right);
\end{equation} 
\begin{equation}
\label{eq43}
y_1 = f_y \left( {x_0 ,y_0 ,z_0 } \right) \equiv f_y \left( {\vec {r}_0 } 
\right);
\end{equation} 
\begin{equation}
\label{eq44}
z_1 = f_z \left( {x_0 ,y_0 ,z_0 } \right) \equiv f_z \left( {\vec {r}_0 } 
\right).
\end{equation} 
Substituting the expressions ((\ref{eq40} and (\ref{eq41} into the right side of equation (\ref{eq38})

$$x_3 = f_x \left( {x_2 ,y_2 ,z_2 } \right) = f_x \left( {x_2 ,f_y \left( 
{\vec {r}_1 } \right),f_z \left( {\vec {r}_1 } \right)} \right) \equiv $$
\begin{equation}
\label{eq45}
\equiv \tilde {f}_x \left( {x_2 ,x_1 ,y_1 ,z_1 } \right),
\end{equation}
and further expressions ((\ref{eq43},(\ref{eq44}) into (\ref{eq45}) we get:

$x_3 = \tilde {f}_x \left( {x_2 ,x_1 ,y_1 ,z_1 } \right) = \tilde {f}_x 
\left( {x_2 ,x_1 ,f_y \left( {\vec {r}_0 } \right),f_z \left( {\vec {r}_0 } 
\right)} \right) \equiv $
\begin{equation}
\label{eq46}
\equiv \tilde {\tilde {f}}_x \left( {x_2 ,x_1 ,x_0 ,y_0 
,z_0 } \right).
\end{equation} 
In order to exclude variables $y_0 ,z_0 $ in (\ref{eq46}) we use the ratio (\ref{eq39}), having substituted expressions for $y_1 ,z_1 $ (\ref{eq44}, \ref{eq45}) and ratio (\ref{eq42}) in it beforehand. 

\[
\begin{array}{l} x_2 = f_x \left( {x_1 ,y_1 ,z_1 } \right) = f_x \left( {x_1 ,f_y \left( {\vec {r}_0 } \right),f_y \left( {\vec {r}_0 } \right)} \right) \equiv \\
\equiv \tilde {f}_x \left( {x_1 ,x_0 ,y_0 ,z_0 } \right); \Rightarrow \end{array}
\]

\begin{equation}
\label{eq47}
\left\{ {{\begin{array}{*{20}c}
 {x_2 = \tilde {f}_x \left( {x_1 ,x_0 ,y_0 ,z_0 } \right);} \hfill \\
 {x_1 = f_x \left( {x_0 ,y_0 ,z_0 } \right).{\begin{array}{*{20}c}
 \hfill \\
\end{array} }{\begin{array}{*{20}c}
 \hfill \\
\end{array} }} \hfill \\
\end{array} }} \right.
\end{equation} 
If the mapping (\ref{eq47}) is biunique relatively to the pair of variables $y_0 
,z_0 $, i.e. if there is a single solution of the set (\ref{eq47}):

\begin{equation}
\label{eq48}
\left\{ {{\begin{array}{*{20}c}
 {y_0 = f_y^{ - 1} \left( {x_2 ,x_1 ,x_0 } \right);} \hfill \\
 {z_0 = f_z^{ - 1} \left( {x_2 ,x_1 ,x_0 } \right),} \hfill \\
\end{array} }} \right.
\end{equation} 
then, substituting $y_0 ,z_0 $ from (\ref{eq48}) into (\ref{eq46}), we will finally receive:

$x_3 = \tilde {\tilde {f}}_x \left( {x_2 ,x_1 ,x_0 ,f_y^{ - 1} \left( {x_2 
,x_1 ,x_0 } \right),f_z^{ - 1} \left( {x_2 ,x_1 ,x_0 } \right)} \right)\equiv$ 
\begin{equation}
\label{eq49}
\equiv F_x \left( {x_2 ,x_1 ,x_0 } \right).
\end{equation} 

If the inverse mapping $(x_1 ,x_2 )$ in $(y_0 ,z_0 )$ for (\ref{eq47}) is not a 
single one, it is necessary to carry out the reasoning, similar to the one 
conducted in the case of two-component model, which leads to the possibility 
of existence of the ``long'' single-component memory in the mapping for the 
$x_n $ component.

Similar calculations and reasoning can be carried out for $N = 4,5,6$ etc., 
and the conclusions will remain the same. It is also obvious that we can 
consider any other component instead of $x_n $ in any situation, what will 
lead only to the change of components indices; it is also possible to consider groups of components, which form any part of the initial component set.

The idea of bringing the set of equations for the multicomponent model to 
one equation (a group of lower equation count) for one of the components 
(group of components) is, as it has been already mentioned, analogous to the 
idea of bringing the system of ordinary first-order differential equations 
to one differential equation (group of quations) of higher order for one of 
the initial (a group of the initial) unknown functions. However there is an 
important difference -- aftereffect, i.e. memory, does not appear in the set 
of differential equations because of the limiting process (the size of pace 
according to time $\Delta t$ tents to zero).

Let us imagine for a moment, hypothetically, a dynamic Universe model as the 
complex nonlinear autonomous system, which started functioning within the 
bounds of a discrete model of the (\ref{eq19}) type in some reasonably distant 
initial moment of time $t_0 $.

Taking into consideration the huge initial number of components of such a 
model and complex, nonlinear character of their interactions, we can assume 
that sufficiently long observation of some limited part of its components 
will show the ``long'' memory, the uncertainties, the absence of repetitions 
(creation of new information) etc. At least, the analysis conducted above 
does not exclude such a possibility, though the realization of it is likely 
to be a rather rare phenomenon in our Universe both in time and space, 
demanding a number of specific circumstances. Our Earth could serve as an 
example of such a realization, having reached a noosphere (the highest for 
today) phase of its development by now.

\section{New paradigms and problems of complex systems mathematical 
description}

Having conducted the afore-mentioned analysis, we made some conclusions, and 
not claiming to make it universal we will briefly dwell on some problems of 
philosophic, conceptual and technical nature, that appear during 
mathematical modeling of real complex systems discussions and problem 
statement. 
\subsection{About the nature of uncertainties and role of action in 
mathematical statement of a problem}

When the attempts to describe the mechanism of the evolutionary development 
of the Universe, which would take into account the practical impossibility 
of an accurate future prediction, are taken the two paradigms collide:
\newcounter{N61a}
\begin{list}{\alph{N61a})}{\usecounter{N61a}}
\item incompleteness of the information on the Universe, including its past, and 
rough character of any model as a result;
\item probabilistic nature of future against the present.
\end{list}

Both paradigms are virtually untestable though.

Indeed, concerning the first paradigm, any information on the system must 
have a material object, which is either a \textit{part} of the system (and cannot contain the full description of it), or an \textit{external} system, interacting with it, i.e. a part of a new fuller system. In this case the interpretation of the process uncertainty is brought to different variations of hidden variables model \cite{bib018,bib105,bib106,bib107,bib108} within the bounds of this paradigm.

The second paradigm virtually comes from the hypothesis of existence of 
multiple, absolutely identical parallel worlds (the quantum ensemble of 
worlds) in every moment of time, when each of them can develop itself 
according to its own probabilistic scenario, but only one of them is 
realized in our world and observed by us \cite{bib018} (the many-world interpretation  was suggested first in \cite{bib109} with the prehistory of it in \cite{bib110}). Thus, according to this paradigm, the real world dynamics is a chain or a sequence of events, having a \textit{random} component of the quantum-mechanical nature. 

However, the notion of an accidental event and probability assumes a 
hypothetical possibility of infinite experiment repeatability under 
identical conditions, and, by the reason of it, the probability theory must 
be considered to be merely one of possible and deliberately approximate 
models of description of uncertainties observed in the world. 

In fact there are no accurate procedures, which would give the opportunity 
to distinguish the ``true'' random sequence of events or quantities from the 
``pseudorandom'' one, i.e. the one similar to the arbitrary, such as 
generated by any suitable determinate chaos model. Really and truly any 
``random'' finite sequence cannot be random because of its finiteness, and 
any ``nonrandom'' finite sequence can be considered to be the one of 
possible and scarce samples of a true \textit{infinite} random sequence. (Here we proceed from the idea, that the notion of infinity is the one of hypotheses, unverifiable on principle, which included as one of the postulates into the 
rigorous theory of sets \cite{bib077}.) 

Moreover, socio-economic phenomena don't repeat themselves \textit{accurately}, and quite low disturbances in real systems can lead to rather big anomalies, which are hard to predict (crises, crashes, bankruptcies and other phenomena of 
critical character, that usually show their individual and unique 
peculiarities).

Both paradigms mentioned above proceed from the assumption that there is a 
notion of system \textit{state} and this notion is the primary and fundamental one. However, repeating the above written, if take into consideration those 
conceptually new things the modern theoretical physics has brought into the 
world, including the relativity theory and relativistic quantum mechanics, 
and be consistent in application of the general system theory, the notions 
of measuring \textit{procedure} and \textit{interaction} between the system and measuring tool, i.e. the result of the \textit{process}, become primary and fundamental. It seems to us that with such statement of a question uncertainty of the state becomes merely a technical problem. Particularly, within the bounds of quantum mechanics uncertainty of the state, i.e. of the quantities characterizing it, is a consequence of certain commutation relations of algebra of \textit{operators} of these quantities \cite{bib052}. 

For justice' sake it is necessary to mention, that such a point of view on 
the fundamental role of action, not the status, was upheld by the prominent 
world and native psychologist and philosopher S. L. Rubinshtein, who is the 
author of the fundamental work ``Fundamentals of General Psychology'' \cite{bib053}, written more than forty years ago, but still actual. A scientist of encyclopedic knowledge, educated in the field of natural sciences, 
mathematics, psychology and philosophy, S. L. Rubinshtein conducted a 
brilliant analysis of historical development of conceptual fundamentals of 
scientific world-view. The authors think that he consciously did not use 
mathematical formalism, realizing that the language of mathematics of 
``states'' and ``functions'' known to him is not appropriate for the level 
and essence of problems, he was solving. 

Economists involved in researches and discussing fundamental problems of 
modern economical theory, use mathematical language carefully or don't use 
it at all (even nowadays, in time of ``informatization'' and 
``computerization''), preferring to bring in their own, new and ex facte 
unusual notions, when the doubts in its adequacy appear. Thus, the notion of 
\textit{coordination} is brought in to characterize the stable socio-economic system state in the monograph written by famous French scientist and practitioner J. Sapir ``Economic theory of heterogeneous systems: an essay on decentralized economies'' \cite{bib049}. It is impossible to bring this notion to such mathematical or physical concepts as equality, identity, equivalence, equilibrium, stationarity etc. This notion should rather be considered to be some kind of a specific characteristic of the non-stationary \textit{action}, which secures stable and steady structural existence for socio-economic system. Here we find implicit ``economical'' arguments for the thesis on the priority of the procedure in description of complex systems dynamics.

Sufficiently persuasive evidences in favour of our positions are present in 
works of the greatest specialist in both classical mathematics and 
mathematical fundamentals of modern quantum theory, academician V. P. Maslov. In 
his latest work, dealing with the mathematical model of the world economical 
crisis of 2008 \cite{bib031}, he clearly shows that the probability theory and the theory of optimization, which form the fundamentals of modern economic 
science, are inadequate as the mathematical toolbox for dynamic description 
of modern economy. On his opinion, the Kolmogorov complexity theory \cite{bib095}, based on the algorithmic approach, should be used as an alternative. 

And, finally, we can't help mentioning the empiriomonism of famous Russian 
politician, economist and thinker A. A. Bogdanov \cite{bib111} and his organizational science -- tectology \cite{bib064, bib065}. His ideas are close to ideas of the general system theory, having anticipated cybernetics, had been wrongly forgotten because of the political motives (both in the West and in Russia) for almost a century. These ideas have actually outstripped their time for century, and only now they start to enter the modern science. His interpretation of organization as the action, which is the fundamental element of the process of functioning in any system, is rather similar to ours and other modern conceptions in philosophical sense. It corroborates the old conception one more time: any new thing or idea is a well-forgotten old one, having been 
pulled out and rediscovered in the ``right'' time and in the ``right'' 
place. Unfortunately he thought about the Einstein relativity theory rather 
critically and wasn't thoroughly aware of quantum physics, arising at that 
time and being beyond his scientific interests.

Thus the conceptions, not necessarily coincident with the traditional ones, 
should form the fundamentals of mathematical modeling of complex systems 
dynamics of any nature. Relativistic quantum mechanics, as it has been 
already mentioned, can serve as one of such sources, however, a certain 
level of caution will be required in this case.
\subsection{About peculiarities, problems and correctness of quantum 
mechanical socio-economic systems modeling}

Most of the researchers who use quantum-mechanical models to explain 
socio-economic phenomena, market dynamics in particular, assume that state 
distribution of the set of its agents (by state strategies are meant) 
conform to Bose-Einstein statistics (e.g. \cite{bib018, bib020, bib021, bib023, bib031, bib032, bib033, bib034}). It means that at one state (one strategy) an arbitrary large amount of agents can coexist. Is it really like that?

If analyze the real behaviour and interrelations of market (or any other 
socio-economic process) participants thoroughly, it is possible to make a 
conclusion, that the equilibrium condition (``equilibrium'' competition) is 
not a fundamental phenomenon, moreover it is a relatively rare one. During 
any kind of interaction in real systems domination relations quickly get 
established, since they are more constructive and stable -- and that is, if 
speaking of quantum-mechanical analogies, rather the Fermi-Dirac statistics 
(only one agent can be at each state).

From the microparticles identity principle and equations of quantum 
mechanics comes a special quantum-mechanical exchange interaction, which is 
implemented in the group of identical particles and put into effect 
according to the "each to every other" principle \cite{bib051, bib052}. However, this principle is of local nature and can get broken, if the size of the system 
considerably exceeds the product of the light speed and the time of 
observing the system (lagging effects). Mechanical transfer of the 
interaction mechanism (according to the principle of intermeshed exchanges) 
to socio-economic systems, where agents play the part of microparticles, and 
the relativistic interaction lagging effect analogues are not necessarily 
connected with light velocity, is not quite competent.

The sufficiently successful explanation of some statistic characteristics 
and dynamic peculiarities of market behavior, derived from 
quantum-mechanical calculations with the help of Bose statistics \cite{bib018}, may not be connected with choosing that very quantum-mechanical model. 
Multicomponent nonlinear models, e.g. (\ref{eq19}), can give rise to rather rich and various scenarios of the dynamic system behavior, even under the 
circumstances of small quantity of varying parameters and variables 
(determinate chaos models \cite{bib056}). Such models can be tangential to equations of quantum mechanics, but let us emphasize that fundamental 
quantum-mechanical principles are present in them and observed in their 
essential peculiarities.

Complex systems are usually synergetic systems with ``long memory'' 
(information on their history), characterized by intensive metabolism 
(constant ``pumping'' of energy and substance) and able to generate new 
information. Formal quantum-mechanical problem statement, pretending to be 
the one to make a detailed ``microscopic'' description of such a complex 
system, can turn out to be inadequate to the processes that really take 
place in the system, although it will reproduce some external peculiarities 
of its behavior. 

By the same reason conservation laws that form the basis of equations of 
physical dynamics and must answer physical processes can have no analogues 
in socio-economic processes. Indeed, in such processes an informational 
component is present (including informational asymmetries of agents \cite{bib112}), transaction costs are possible (the ``fifth'' market \cite{bib113}) and memory occurs (institutions, mentality \cite{bib114, bib115}), the energy and substance receipt and dissipation take place, other types of ``rough'' and ``delicate'' interaction between the environment and the past are also possible. 

Real non-linear interactions in the multi-component socio-economic system 
can change the relations between agents and generate a complex dynamics in 
the way, that traditional analysis, conducted according to the scheme 
``structure-state-interaction-dynamics'' would hardly explain anything 
concerning dynamic system behaviour (synergetic effect, aftereffect, ``long 
memory'', threshold phenomena, conditioned by weak interactions with the 
environment etc.). On the other hand structureless ``field'' approaches, 
based on the ideas of the quantum field theory (the unified field theory), 
if developed, would possibly turn out to be not productive as well. 

On our opinion only measured, discrete by definition, data series, 
characterizing dynamic change of system state during quite long time period 
$T$, can serve as the source of information about complex system. 

In this case that problem statement becomes acceptable and reasonable, where 
the approximate prediction of system behaviour, its informative 
characteristics and algorithm design are considered. Such a statement is 
typical for the new scientific direction in socio-economic processes -- data 
analysis (developed since 1990) \cite{bib116, bib117}. 

Concerning time irreversibility and discrecity, it can be added that time 
irreversibility must be considered as an experimentally found within the 
bounds of its application, fact. Time characterizes duration of procedures, 
processes, phenomena, i.e. the duration of actions, and can be determined 
only with the help of various actions. Minimal time interval is actually 
determined by the observed action of minimal duration. However, according to 
the special and general relativity theories this notion (i.e. the idea of 
duration) must be considered as the one, which is relative, local in time 
and space and depending on the coordinate system [\cite{bib118}]. 

In theoretical physics, energy and momentum (angular momentum) conservation laws are considered as fundamental consequences of homogeneity of our time and space (space isotropy) [\cite{bib084}]. Hypotheses on their continuity are the convenient, but not necessary component for receipt of respective laws of conservations. For example the energy conservation law can be considered as the universal postulated technology of detection of new (or already known) interactions and types of energy and substance transformation in physical systems. Thus the new elementary particle neutrino was discovered as the consequence of formally observed failure of conservation laws during experiments on $\beta $-decay of radioactive elements. (The weak interaction connected with 
neutrino was so ``weak'' that this particle can fly through the Sun and 
experience no collision.) The other example is Einstein's ratio of energy 
and mass $E = mc^2$, which tied physical quantities, considered to be 
heterogeneous before that.

On our opinion, during mathematical modeling of complex systems laws of 
conservation of various quantities, time irreversibility or reversibility, 
its discrecity and continuity, homo- or heterogeneity etc. must be 
considered as the properties of this very mathematical model, first of all 
appreciating the level of its adaptation to the description of \textit{real} properties 
and \textit{real} system dynamics, the history of which must be considered as unique experiment data, not always possible to repeat. The level of system adequacy to the processes investigated, maximum possible predictability and practical significance must serve as the basic criteria of the model. 

\section{Conclusion}

To resume we will briefly formulate new paradigms and main conceptual 
statements in  complex systems modeling, which come from the analysis we 
conducted.
\newcounter{concl1}
\begin{itemize}
\item Priority of the measuring (observing, action, interaction) procedure against its result;
\item Unoriginality and approximate nature of notions of ``system state'' and 
``immediate values of quantities'' as characteristics of this state;
\item Finite length and unremovable influence of any measuring procedure, including computer prediction (indirect measurement), the state and future behaviour of the system;
\item Uncertainty principle and its fundamental connection with the duration of 
the measuring procedure; 
\item Discreteness of time, space and any other quantity, connected with the notion of state and system dynamics;
\item Aftereffect (memory) as the fundamental quality of any complex dynamical 
system;
\item Refusal of infinity as the conceptual notion;
\item Time irreversibility;
\item Openness;
\item Hierarchy;
\item Emergence.
\end{itemize}

Some of the afore-mentioned positions coincide with positions of the general 
system theory, what is not strange from the one side, and allows 
interpreting our analysis as the physical quantum-mechanical substantiation 
of system conceptions in modeling complex systems from the other one \cite{bib022}.

In this analysis and conclusions facts and postulates of relativistic 
quantum physics and experience of observing and researching real 
socio-economic systems are considerably used, which gives us the reason to 
relate this work to the new direction in physical economics, declared in the 
name -- relativistic quantum econophysics.

We have begun specific research and development on realization of the above-listed conceptions in modeling and prediction of socio-economic processes, based on the observation data (history) of relevant time series \cite{bib119, bib120,bib121}. One of the prediction technologies is based on the use of complex Markov chains (Markov chains with memory), is implemented in Matlab 6.5 environment and is currently being tested on the stock fund indexes and exchange rates data.

\bibliography{comments,refs}

\end{document}